\documentclass[preprint,showpacs,floatfix]{revtex4-1}
\usepackage[dvips]{graphicx}
\usepackage{epsfig}
\usepackage{color}
\usepackage[normalem]{ulem}
\usepackage{amsmath,amsfonts,amssymb,bm}
\setcitestyle{numbers,square}
\usepackage{verbatim}
\usepackage[capitalize]{cleveref}
\usepackage{color,soul}
\usepackage{sans}
\usepackage[labelformat=simple]{subfig}

\UseRawInputEncoding

\begin{document}
\title{Maximum propagation speed and Cherenkov effect in optical phonon transport through   periodic molecular chains}

\author{Alexander L. Burin, Igor V. Parshin and Igor V. Rubtsov}
\affiliation{Department of Chemistry, Tulane University. New Orleans, LA 70118, USA}

\date{\today}
\begin{abstract}
Optical phonons serve as the fast and efficient carriers of  energy across  periodic polymers due to their delocalization, large group velocity because of covalent bonding and large energy quantum compared to that for acoustic phonons, as it was observed in a number of recent measurements in different oligomers. However, this transport is dramatically sensitive to anharmonic  interactions, including  the unavoidable interaction with acoustic phonons responsible for the transport decoherence, suppressing  ballistic transport at long distances. Here we show that this decoherence is substantially suppressed  if the group  velocity of optical phonons is less than the sound velocity of acoustic phonons; otherwise ballistic transport is substantially suppressed by a Cherenkov's like emission of acoustic phonons. This conclusion is justified  considering energy and momentum conservation during phonon absorption or emission and supported by the numerical  evaluation of lifetimes of the optical phonons. It is also consistent with the recent experimental investigations of ballistic optical phonon transport in oligomers with minor exception of relatively short oligophenylenes.  
\end{abstract}
\maketitle

\section{Introduction}

Vibrational energy relaxation and transport are generally responsible for the  chemical energy balance in molecular systems \cite{AbeScience07,15LeitnerReview,
PandeyLeitner2016ThermSign} and they are relevant  for modern molecular devices capable to transfer heat at nanoscales  \cite{Cui2019SingleMoleculeThermCond,Abe03,
15LeitnerReview,SegalReview2016}.  Within this settings,  chemistry at a distance may be accomplished similarly to that for the charge transport in DNA \cite{Barton98ChemAtDist} transferring the energy between two separated reaction centers using the oligomer bridge.   Therefore, efficient vibrational energy transport  is important to several fields spanning from sustainable energy to biomedicine to thermal management \cite{Zhu19AdvMatVibrEnTr}.  Intramolecular energy transport mediated by optical phonons can be fast and efficient in periodic polymers \cite{Troe04BallFirstExp,DlottScience07,AbeScience07,SegalNitzan03,SegalReview2016,Asegun08Polyethilenetransport, Rubtsov12Acc,ab15AccountsIgor,ab19IgorReview,QuantizTranspPhDviraRev22,
Cui2019SingleMoleculeThermCond}. Indeed,  periodicity results in   delocalization of normal vibrational modes that is required for the ballistic transport, while covalent bonds between atoms lead to high transport velocities compared to molecular materials. 

Most of the present studies treat thermal energy transport \cite{Troe04BallFirstExp,DlottScience07,SegalNitzan03,
SegalReview2016,Asegun08Polyethilenetransport, QuantizTranspPhDviraRev22,
Cui2019SingleMoleculeThermCond}  that is mediated by acoustic phonons at small deviations from equilibrium occurring at room temperature and below. Under these conditions  optical phonons possess a high activation energy, and therefore are not excited within that temperature range. However, if the heat is released in chemical reactions accompanied by substantial changes in electronic states, then these changes necessarily involve excitations of    polar optical phonons  \cite{BIXON1993ElTransfReorg,
Hsu2020PCCPReorganizationPhononsRev}. Similarly,  an excitation of high frequency optical vibrations should accompany  electronic excitations  by laser or natural sunlight e. g. in the course of photosynthesis \cite{Stock2022EnTranspOptProt,Stock2019EnTrPOthProt,Rich2020JPCLett,
Champion2005VibrEnRelOpt}. Under all mentioned conditions the fast energy transport mediated by optical phonons can determine the energy excess transport from the reaction and/or photo-excitation center.   The possibility of the fast and efficient optical phonon transport with the transport speed reaching tens of thousands meters per second has been demonstrated in two-dimensional infrared (2DIR) studies of the optical phonon transport via several oligomeric chains.\cite{Rubtsov2012pphynilultrfast,
Rubtsov2009Accounts2DIR,ab15ballistictranspexp,
ab19layla,ab19IgorReview}

The most efficient optical phonon  transport is realized in the ballistic regime where the wavepacket formed at the end of the polymer chain propagates across the chain to its opposite end in an entirely coherent manner. The characteristic group velocity for such transport is determined by the optical phonon bandwidth \cite{ab16jpcPEGs}. This transport is limited, however, by  the full Anderson localization of all  vibrational modes due to possible  chain disordering \cite{Lebowitz67,Abrahams1979ScThLoc}  as well as by   relaxation and decoherence due to anharmonic interactions. 


Decoherence and relaxation result in the loss of quantum coherence for the propagating phonon. Decoherence processes conserve  a number of optical phonons within the band. They  are possible due to the interaction of an optical phonon with low energy acoustic phonons forming the gapless bands with a minimum energy equal to zero  \cite{landau1986theory,PolimVibr1994Book,
Chico2006VibrCarbNanot,
NanoTubeVibr20071,ab20Transv,ab15jcp}.  The relaxation is accompanied by disappearence of optical phonons from their initial band. 

Experimental investigations of optical phonon transport in various oligomers \cite{ab15JPCExpDec,ab16jpcPEGs} show that decoherence emerges earlier than relaxation or Anderson localization at least at room temperature expressing itself as the transition from the ballistic to the diffusive transport at molecular lengths exceeding a coherence length. Decoherence  can take place faster compared to the relaxation   because relaxation necessarily involves a  weak overlap of vibrational modes belonging to  different  phonon bands or solvent, while  decoherence is due to the interaction with acoustic phonons usually delocalized through the whole molecule. 

The latter circumstance determines the focus of the present work on the restrictions of ballistic energy transport due to  decoherence caused by an interaction of a propagating optical phonon with acoustic phonons. One should notice that these restrictions can be important also for thermal energy transport mediated by acoustic phonons belonging to different  acoustic  bands of one-dimensional oligomers \cite{landau1986theory,Chico2006VibrCarbNanot}.

The simplest anharmonic interaction  leading to decoherence is an absorption or emission of an acoustic phonon by a propagating optical phonon accompanied by its transition between two states within its energy band (cf Ref. \cite{Leitner2001ThirdOrdAnh}). Then it is natural to expect that  decoherence  is sensitive to the relationship between the speed of sound $c$ and the transport velocity of the optical phonon $v_{tr}$   similarly to the celebrated Cherenkov's effect \cite{Jackson1998TextBookCheren}, where  the emission of light by the moving particle  emerges if a  particle  speed exceeds  the  speed of light in that medium. The Cherenkov's  emission of high-frequency acoustic phonons was earlier considered for drifting
electrons in a quantum well (Si/SiGe/Si device) \cite{Kornienko2000CherenEmissAcPh}, quantum well
heterostructures  \cite{Komirenko2001AmplCheren},  graphene  \cite{Zhao2013CerenGraph} and transition metal dichalcogenides \cite{Kubakaddi2017Cheren}.  
Optical phonons should emit acoustic phonons   similarly to electrons.  Consequently,  one can expect a strong decoherence of optical phonons with the propagation speed exceeding the speed of sound and substantially suppressed decoherence in the opposite case. As a result, the velocity of efficient energy transport via optical bands is restricted from the top by the speed of sound. This expectation is essentially confirmed by our numerical results reported in the present work.

The paper is organized as following. In Sec. \ref{sec:ModHv} the model for the optical phonon transport is introduced and the elementary process of the emission of acoustic  phonons is considered from the perspective of a simultaneous conservation of energy and momentum (wavevector), which are possible only for the optical phonon group velocity exceeding the velocity  of acoustic phonons, i. e. speed of  sound. In Sec. \ref{Sec:Diag} the  lack of decoherence or its emergence depending whether Cherenkov's emission is forbidden or allowed is demonstrated numerically. The lack of decoherence is interpreted in terms of the localization within the Hilbert space of all possible  states with defined populations of normal modes  \cite{LoganWolynes90} that we can refer as {\it harmonic states}, i. e. eigenstates of the system without anharmonic interactions. One should notice that this localization emerges in the momentum space in contrast to the Anderson localization  \cite{Anderson58}   occurring in the coordinate space since all normal modes in the periodic system possess a certain momentum and they are delocalized in space through the whole molecule. Consequently, the localization in the momentum space suggests the  full delocalization in the coordinate space. 

Based on the level statistics, we show that decoherence is suppressed in the absence of Cherenkov's emission in a wide range of temperature above zero temperature.  In Sec. \ref{sec:Exp} our expectations are compared to the available experimental data for perfluoroalkane \cite{ab15JPCExpDec}, alkane \cite{ab15ballistictranspexp,ab15jpc}, PEG \cite{ab16jpcPEGs} and oligo(p-phenylene)  \cite{Rubtsov2012pphynilultrfast} oligomers. The results are summarized  in Sec. \ref{sec:Concl}. 

We restrict our consideration to acoustic modes with  sound-like spectra Eq. (\ref{eq:AcPhNN}),  which are   longitudinal and torsional modes \cite{landau1986theory,PolimVibr1994Book}. The spectrum of transverse vibrational modes is different \cite{ab20Transv} and decoherence due to transverse modes needs a separate consideration. We also ignore disordering assuming that  Anderson localization length substantially exceeds the decoherence length. The latter assumption seems to be valid at least in perfluoroalkanes  \cite{ab14PerFluoroAlkExp} and alkanes \cite{ab15ballistictranspexp} and  it is supported by an advanced  numerical studies of energy transport in disordered chains \cite{Allen1998DisordVibrWavePack}. We also ignore  anharmonic interactions of acoustic phonons that can lead to  decoherence  \cite{Leitner2001ThirdOrdAnh}  and  soliton formation \cite{Chetverikov2006Soliton}, since both phenomena emerge slower and at longer length  scale compared to the Cherenkov's emission, because they involve  higher order anharmonic interactions. 

\section{Model and Cherenkov Effect}
\label{sec:ModHv}

\subsection{Optical phonon interacting with acoustic phonons}

For the sake of simplicity we consider periodic polymer chains. Although real chains are not periodic the propagation of phonon should not be sensitive to the boundary conditions if the chain length is long enough compared to the coherence  length.  In that regime the boundary conditions are not very significant, while in the opposite regime   the maximum transport efficiency is attained. Thus our estimate for the coherence length should be relevant independent of the boundary conditions. The use of periodic model  substantially  simplifies numerical studies because of the quasi-momentum conservation   (see   Sec.  \ref{Sec:Diag} and Ref.  \cite{ab19FPU}) that permits us to investigate numerically very long chains up to $12$ unit cells. 

The  optical phonon band for the lattice with the period $a$ is characterized by the spectrum $\omega_{opt}(k)$ representing the frequency dependence on the wavevector $k$ that should be periodic with the period $2\pi/a$. For estimates we will use the simplest tight binding model of the optical phonon transport by means of hops between nearest neighbors with the amplitude $\Delta/4$, where $\Delta$ stands for the bandwidth. In this model  the energy spectrum can be represented as 
\begin{eqnarray}
\omega_{opt}(k) = \omega_{0}+ \frac{\Delta (1-\cos(ka))}{2},  
\label{eq:OptPhNN}
\end{eqnarray}
where $\omega_{0}$ is the optical phonon bandgap (we assumed $0<\Delta\ll \omega_{0}$). The group velocity can be expressed  as the derivative of the frequency Eq. (\ref{eq:OptPhNN})   with respect to the wavevector. This yields 
\begin{eqnarray}
v_{opt}(k) =\frac{d\omega_{opt}}{dk}  =v_{max}\sin(ka), ~v_{max}=\frac{\Delta a}{2}.  
\label{eq:OptPhgv}
\end{eqnarray}
Here $v_{max}$ is the maximum group velocity of the optical phonon. 

One can describe acoustic phonons similarly by the gapless dispersion law $\omega_{ac}(k)$, suggesting $\omega_{ac}(0)=0$. Using a similar tight-binding approximation one can model  the acoustic phonon spectrum as \cite{ab19FPU} 
\begin{eqnarray}
\omega_{ac}(k) = \frac{2c}{a} |\sin(ka/2)|,  
\label{eq:AcPhNN}
\end{eqnarray}
where $c$ is the velocity of sound. The group velocity for the acoustic phonons can be expressed similarly to Eq. (\ref{eq:OptPhgv}) as 
\begin{eqnarray}
v_{ac}(k) =\frac{d\omega_{ac}}{dk}  =c\cos(ka/2). 
\label{eq:ActPhgv}
\end{eqnarray}

\begin{figure}
\includegraphics[scale=0.5]{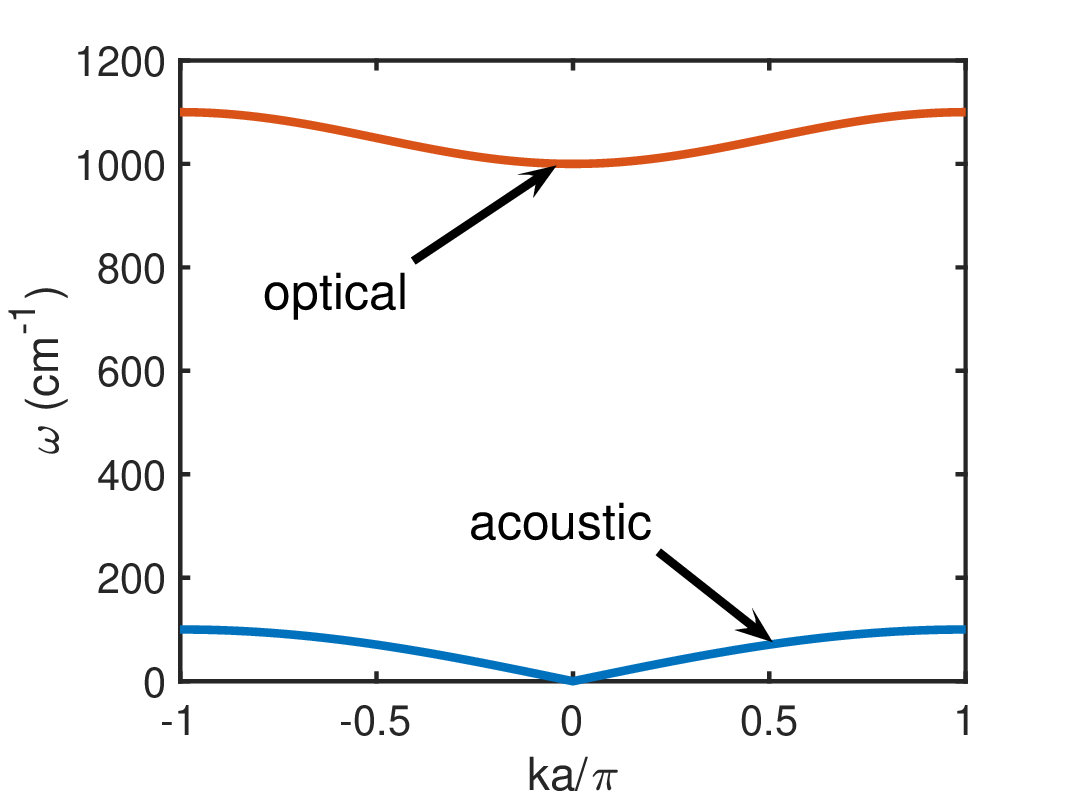} 
\caption{Acoustic and optical phonon spectra for the parameters $\omega_{0}=1000$cm$^{-1}$, $\Delta=100$cm$^{-1}$, $a=3$\AA, $c=5.57\cdot 10^{3}$m$/$s.}
\label{fig:Spectr}
\end{figure}

The representative spectra for acoustic and optical phonons are shown in Fig. \ref{fig:Spectr}. For the parameters used  there the maximum optical phonon group velocity $v_{max}=\Delta a/2$ is equal to the speed of sound $c$. 

In our consideration we use the same parameters for the optical phonon band as  defined in captions to Fig. \ref{fig:Spectr}, while the width of the acoustic band and, correspondingly, the speed of sound varies to probe different  regimes for the Cherenkov's emission. The position of the band and the bandwidth are quite typical for organic polymers  \cite{ab15JPCExpDec,ab15ballistictranspexp,ab15jpc,
ab16jpcPEGs}.

The Hamiltonian of non-interacting acoustic and optical phonons can be expressed as 
\begin{eqnarray}
\widehat{H}_{0}=\sum_{k}\hbar\left(\omega_{ac}(k)\widehat{a}_{k}^{\dagger}\widehat{a}_{k}+\omega_{opt}(k)\widehat{b}_{k}^{\dagger}\widehat{b}_{k}\right),  
\label{eq:Ham0}
\end{eqnarray}
where the Bose-operators $\widehat{a}$ ($\widehat{a}^{\dagger}$) or $\widehat{b}$ ($\widehat{b}^{\dagger}$) represent annihilation (creation) operators of acoustic or optical phonons, respectively.  The wavevector $k$ takes $N$ different values $0$, $\pm 2\pi/L$, ... , where $L$ stands for the chain length and $N=L/a$ is the number of unit cells.

Anharmonic interactions can be treated as a perturbation to the harmonic problem Eq. (\ref{eq:Ham0})  since they are typically  smaller by the factor of the order of $0.1$. This factor  expresses  the small ratio of atomic displacement and interatomic distance. Consequently, we consider  the third order   anharmonic interaction that is of the lowest  order in atomic displacements. In that order the only processes capable to conserve energy are determined by  absorption or emission of acoustic phonons by optical phonons  as  (remember, that we ignore interband relaxation processes representing internal vibrational relaxation)
\begin{eqnarray}
\widehat{V}_{anh}=\frac{1}{\sqrt{N}}\sum_{q,k_{1},k_{2}}\left[ V(q, k_{1}, k_{2})\widehat{b}_{k_{1}}^{\dagger}\widehat{b}_{k_{2}}\widehat{a}_{q}^{\dagger}  + H. C.\right]\delta_{*}(L(k_{1}+q-k_{2})/(2\pi)),
\nonumber\\
\delta_{*}(n)=\sum_{k=-\infty}^{+\infty}\delta_{n, kN},
\label{eq:AnhIntGen}
\end{eqnarray}
where $V(q, k_{1}, k_{2})$ is the interaction constant, $\delta_{n,m}$ is the Kronecker symbol, wavevectors $k_{1}$ and $k_{2}$ characterize optical phonons and $q$  designates wavevectors of acoustic phonons in interaction terms.
The wavevector conservation with the accuracy to the inverse lattice period $2\pi/a$ is expressed by the generalized Kronecker symbol $\delta_{*}$.  

The interaction with acoustic phonons is proportional to the gradient of displacement $qu_{q}$ in the long wavelength limit $q\rightarrow 0$ since identical displacements of all atoms does not change energy. Since $u(q) \propto (\widehat{a}_{q}^{\dagger}+\widehat{a}_{q})/\sqrt{\omega_{ac}(k)}$ one has $V(q, k_{1}, k_{2})\propto \sqrt{q}$ in the long wavelength limit.

For the numerical studies in Sec. III we take the interaction in the simplest local form 
\begin{eqnarray}
\widehat{V}_{loc}=\frac{V_{0}}{a}\sum_{n=1}^{N}\widehat{b}_{n}^{\dagger}\widehat{b}_{n}(\widehat{u}_{n+1}-\widehat{u}_{n-1}),
\label{eq:ModAnh}
\end{eqnarray}
where $V_{0}$ is the interaction constant, $\widehat{b}_{n}$($\widehat{b}_{n}^{\dagger}$)  are  the optical phonon  annihilation  (creation) operators in the coordinate representation.The operator  $\widehat{u}_{n}$ is the coordinate representation of the displacement operators $u_{q}$ for the  acoustic phonons, defined as 
\begin{eqnarray}
\widehat{u}_{n}=\frac{1}{\sqrt{N}}\sum_{q}u_{q}e^{-iqn}, ~\widehat{u}_{q}=\sqrt{\frac{\hbar}{2M\omega_{q}}}\left(\widehat{a}_{q}^{\dagger} +\widehat{a}_{-q}\right),
\label{eq:AcPh}
\end{eqnarray}
where $M$ is the mass of the unit cell. 
The coordinate $u_{n}$ in Eq. (\ref{eq:ModAnh}) is a periodic function of the position $n$ satisfying $u_{0}=u_{N}$ and $u_{N+1}=u_{1}$. 

The interaction Eq. (\ref{eq:ModAnh}) can be expressed in the general form of the third order anharmonic interaction  Eq. (\ref{eq:AnhIntGen}) using wavevector  representatios for optical and acoustic phonons. Then the interaction constants in Eq. (\ref{eq:AnhIntGen}) are defined as 
\begin{eqnarray}
V(q, k_{1}, k_{2}) = 2iV_{0}\sin(qa)\sqrt{\frac{\hbar}{M\omega_{ac}(q)a^2}}
\nonumber\\
=iV_{3}{\rm sign}(qa)\cos(qa/2)\sqrt{2|\sin(qa/2)|}, ~ V_{3}=\eta V_{0},  ~ \eta=\sqrt{\frac{\hbar }{Mca}},
\label{eq:AnhIntC}
\end{eqnarray}
where the factor $\eta \sim 0.1$ is determined by the ratio of a typical root mean squared atomic displacement $\sqrt{\hbar/(Mc/a)}$ and a  chain period $a$, which is comparable to an interatomic distance. This factor  expresses  relative weakness of an anharmonic interaction compared to a harmonic one. 

The full Hamiltonian used in our analytical and numerical considerations below can be expressed as 
\begin{eqnarray}
\widehat{H}=\widehat{H}_{0}+\widehat{V}_{loc}, 
\label{eq:FullH}
\end{eqnarray}
with the harmonic part $\widehat{H}_{0}$ defined in Eq. (\ref{eq:Ham0}) and anharmonic part $\widehat{V}_{loc}$ defined by Eq. (\ref{eq:ModAnh}). 

The model under consideration does not include directly higher order anharmonic interactions that can be significant for the optical phonon decoherence, particularly in the regime where the acoustic phonon absorption or emission are substantially suppressed. However, those interactions are virtually generated by the third order anharmonic interactions in higher orders of pertubation theory (see Ref. \cite{ab19FPU}, Sec. 3.3). Therefore, in spite of the absence of direct high order anharmonic interactions in our model, our consideration remains valid at least qualitatively, since  generated interactions are of the same order of magnitude with respect to the atomic displacement as the original ones. 

\subsection{Emergence of acoustic photon  absorption or emission: Cherenkov effect.}
\label{subsec:Cher}

An optical  phonon with the wavevector  $k$ can emit  or absorb an acoustic phonon with the wavevector $q$ transferring to the optical phonon state with the wavevectors $k\mp q$ belonging to the continuum of states due to the quasi-momentum conservation.  A decay to the continuum can be described within   the framework of the Fermi Golden rule.  Using the  interaction in Eq. (\ref{eq:AnhIntC}) we express the emission and absorption rates $W_{em}$, $W_{abs}$ as (cf. \cite{Leitner2001ThirdOrdAnh})
\begin{eqnarray}
W_{em}=4\pi V_{3}^2 \frac{a}{2\pi}\int_{-\pi/a}^{\pi/a}dk' \cos((k-k')a/2)^2|\sin((k-k')a/2)|
\nonumber\\
\times\delta\left(\omega_{opt}(k)-\omega_{opt}(k')-\omega_{ac}(k-k')\right)(1+\nu_{ac}(k-k')), 
\nonumber\\
W_{abs}=4\pi V_{3}^2 \frac{a}{2\pi}\int_{-\pi/a}^{\pi/a}dk' \cos((k-k')a/2)^2|\sin((k-k')a/2)|
\nonumber\\
\times\delta\left(\omega_{opt}(k)-\omega_{opt}(k')+\omega_{ac}(k-k')\right)\nu_{ac}(k-k'), ~\nu_{ac}(q)=\frac{1}{e^{\frac{\hbar\omega_{ac}(q)}{k_{B}T}}-1},  
\label{eq:FGr}
\end{eqnarray} 
where $\nu_{ac}(q)$ represents the average number of acoustic phonons with the wavevector $q$. 

\begin{figure}
\includegraphics[scale=0.5]{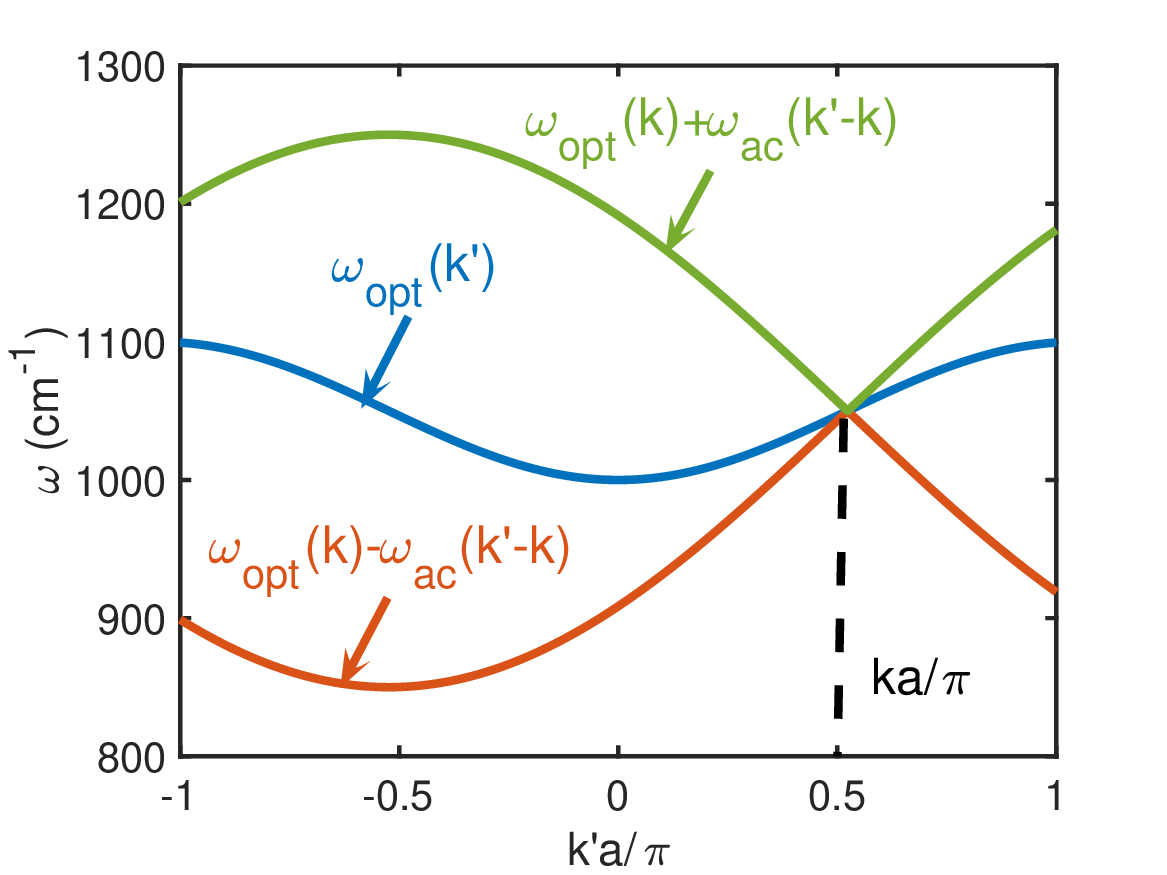} 
\caption{Absence of energy conserving  absorption or emission processes in the case of Eq. (\ref{eq:CherEmisN}). All parameters are  as in Fig. \ref{fig:Spectr}, but the speed of sound is twice larger   ($c=2v_{max}$).}
\label{fig:Forb}
\end{figure}

\begin{figure}
\includegraphics[scale=0.5]{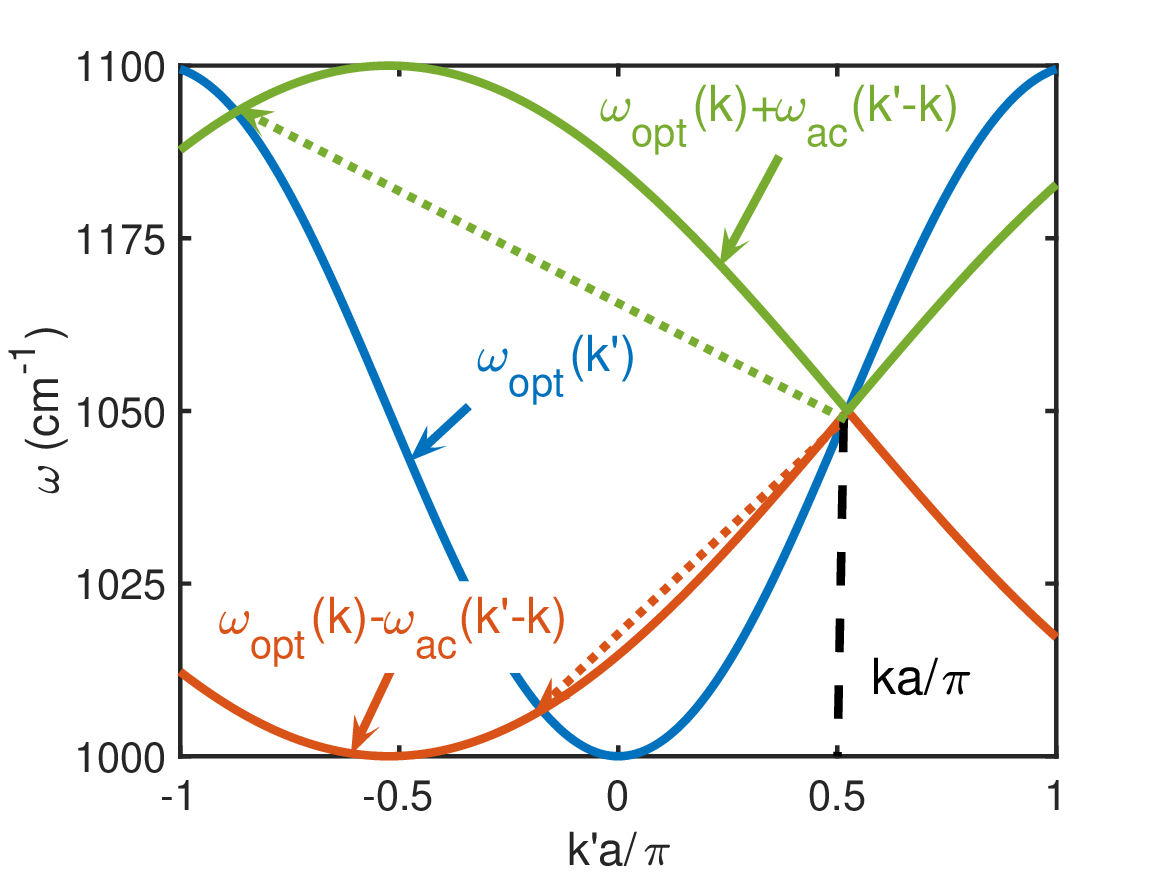} 
\caption{Allowed absorption or emission processes in the case opposite to  Eq. (\ref{eq:CherEmisN}) with parameters as in Fig. \ref{fig:Spectr}, except for the speed of sound  $c=v_{max}/2$. The wavevector $k$ is chosen as in Fig. \ref{fig:Forb}. The downwards or upwards arrows towards intersections show the changes of the wavevector for emission or absorption transitions.}
\label{fig:All}
\end{figure}

According to  Eq. (\ref{eq:FGr}) emission (A) and  absorption  (B) must satisfy the  energy conservation in the forms  
\begin{eqnarray}
\omega_{opt}(k)-\omega_{ac}(k-k')=\omega_{opt}(k'), ~ (A)
\nonumber\\
\omega_{opt}(k)+\omega_{ac}(k-k')=\omega_{opt}(k'), ~ (B).
\label{eq:encons}
\end{eqnarray}
 In Figs. \ref{fig:Forb}, \ref{fig:All} frequencies  $\omega_{\mp}(k')=\omega_{opt}(k)\mp \omega_{ac}(k-k')$ and $\omega_{opt}(k')$ are plotted together for the model with nearest neighbor couplings, Eq. (\ref{eq:OptPhNN}) and Eq. (\ref{eq:AcPhNN}), and   different relationships of the speed of sound  and the maximum velocity of optical phonons.  Their intersections (if any) determine the allowed absorption or emission processes in accordance with Eq. (\ref{eq:encons}) for $k=\pi/(2a)$.  

For  the model with nearest neighbor coupling,  Eq. (\ref{eq:encons}) needed to produce a nonzero emission rate Eq. (\ref{eq:FGr})  can be further simplified as 
\begin{eqnarray}
\sin\left(\frac{a(k+k')}{2}\right) = \pm \frac{c}{v_{max}}{\rm sign}\left(\sin\left(\frac{a(k-k')}{2}\right)\right).
\label{eq:enconsNN}
\end{eqnarray}
with the positive sign for absorption and negative sign for emission. 

This equation has no solutions if the maximum group velocity of optical phonons, $v_{max} = a\Delta/2$ Eq. (\ref{eq:OptPhgv}),  is less then the speed of sound $c$. Thus, if the criterion for the Cherenkov's emission is not satisfied, i. e. 
\begin{eqnarray}
v_{max}=\frac{a\Delta}{2} <c,  
\label{eq:CherEmisN}
\end{eqnarray}
one can expect no emission. The lack of solutions of Eq. (\ref{eq:enconsNN}) is illustrated in Fig.  \ref{fig:Forb}, where Eq. (\ref{eq:encons}) is represented  graphically,  for $v_{max}=c/2$. 


In the opposite case $v_{max}>c$ (see Fig. \ref{fig:All} for   $v_{max}=2c$) the non-trivial ($k'\neq k$) solution of Eq.  (\ref{eq:enconsNN}) always exists, at least either for emission  or for absorption. If $v(k)>c$ then both emission and absorption are allowed as shown  in  Fig. \ref{fig:All}, while for the bottom or the top of the optical band where $v(k)<c$ only absorption or emission are allowed, respectively. If only absorption is allowed then the phonon state has no decay channels at zero temperature. 

The consideration above is developed for the direct band $\Delta>0$. Its generalization to the indirect band $\Delta<0$ is straightforward and can be made by a simultaneous shift of optical phonon wavevectors by $\pi/a$ with the change in the sign of the bandwidth parameter. Consequently, all our results about the absorption or emission of acoustic phonons remain valid.

If absorption or emission are allowed, then the  integrals in Eq. (\ref{eq:FGr}) can be evaluated as (the acostic phonon wavevector $q$ is replaced with the difference of optical phonon wavectors $k_{1}-k_{2}$ due to the wavevector  quasi-conservation, Eq. (\ref{eq:AnhIntGen}))
\begin{eqnarray}
W_{em}=\frac{4V_{3}^2}{\Delta}\frac{1}{1-\exp\left[-\frac{2\hbar c|\sin(a|k|-\varphi)|}{ak_{B}T}\right]}
\frac{\cos(a|k|-\varphi)^2}{\sqrt{1-c^2/v_{max}^2}}, ~ a|k|>\varphi, 
\nonumber\\
W_{abs}=\frac{4V_{3}^2}{\Delta}\frac{1}{\exp\left[\frac{2\hbar c|\sin(\varphi-a|k|)|}{ak_{B}T}-1\right]}
\frac{\cos(ak+\varphi)^2}{\sqrt{1-c^2/v_{max}^2}}, ~ a|k| < \pi-\varphi,
\nonumber\\
\varphi=\sin^{-1} \left[\frac{c}{v_{max}}\right]
\label{eq:FGRAns}
\end{eqnarray}
Replacing the cosine squared factor with its mean value $1/2$ and assuming $c<v_{max}$ we estimate a typical decoherence rate at zero temperature as $r_{d}\approx V_{3}^2/\Delta$. Using typical parameters $\Delta \sim 100$cm$^{-1}$ and $V_{3}\sim 20$cm$^{-1}$ (see Sec. \ref{subsub:V0.2}) we obtain $r_{d} \approx 1$ps$^{-1}$. At longer times  then $r_{d}^{-1}$, ballistic transport is substantially  suppressed. 


The models for acoustic and optical phonon spectra (Eqs. (\ref{eq:OptPhNN}) and (\ref{eq:AcPhNN})) represent the first terms in the Fourier expansions of vibrational frequencies as periodic functions of the wavevector that are associated with  the nearest neighbor coupling. Other  expansion terms can modify both spectra and affect the Cherenkov's emission criterion Eq. (\ref{eq:CherEmisN}).  Yet these modifications should be relatively weak, since they are due to coupling of next to nearest and stronger separated neighbors. Consequently, they  should not strongly modify the criterion of the absence of Cherenkov's emission.

If the molecule is immersed in a solvent, the acoustic phonons of the solvent (medium) can also contribute to the Cherenkov type scattering of the optical phonons of the molecule. For that to occur, one should similarly satisfy the energy conservation laws 
\begin{eqnarray}
\omega_{opt}(k)=\omega_{opt}(k')+\omega_{ac,s}(k-k',\mathbf{q}_{\perp}), ~ (A)
\nonumber\\
 \omega_{opt}(k)=\omega_{opt}(k')-\omega_{ac,s}(k-k', \mathbf{q}_{\perp}), ~ (B)
\label{eq:enconsSolv}
\end{eqnarray}
modified by using the solvent acoustic wave spectrum $\omega_{ac,s}(k-k', \mathbf{q}_{\perp})$ with an arbitrary transverse wavevector component $\mathbf{q}_{\perp}$  perpendicular to the molecular axis. If $v_{max}<c$ then, similarly to the previous consideration, one can expect $\omega_{opt}(k)-\omega_{opt}(k')<\omega_{ac,s}(k-k', 0)<\omega_{ac,s}(k-k',\mathbf{q}_{\perp})$ and, consequently, no Cherenkov's emission.   In a certain extent an infinite  molecule behaves  similarly to the cylindrical waveguide that functions when it has a dielectric constant larger than that of the environment and, consequently, the internal speed of light smaller than that in the environment \cite{LLElectrMed}. 

In the opposite case of $v_{max}>c$ both absorption and emission are possible at least for $q_{\perp}=0$ and, consequently, in general.  
Yet the decoherence rate due to the interaction with solvent should be much smaller than that due to intramolecular interaction estimated as $1$ ps$^{-1}$ since the intramolecular interactions are determined by covalent bonds, while intermolecular interactions are due to dispersive forces. Consequently, one can expect one or two orders of magnitude slower decoherence that can be seen only in very long molecules.

\subsection{Higher order processes}
\label{subsec:2HighOrd}

The above consideration is limited to single phonon emissions or absorptions, which are completely forbidden if the maximum optical phonon group velocity is less than the speed of sound of acoustic phonons.  It is, of course, interesting to understand whether the processes induced by higher order anharmonic interactions are permitted in this regime.

The answer to this question depends on the temperature. At zero temperature the only multiphonon emission processes that can take place must satisfy the energy conservation law in the form 
\begin{eqnarray}
\omega_{opt}(k)-\omega_{opt}(k-q_{1}-q_{2}-...-q_{n})= \omega_{ac}(q_{1})+\omega_{ac}(q_{2})+...+\omega_{ac}(q_{n})
\label{eq:encons-nph}
\end{eqnarray}
generalizing Eq. (\ref{eq:encons}) to the $n$-phonon process. It turns out that in the regime of forbidden Cherenkov emission ($\omega_{opt}(k)-\omega_{opt}(k-q)<\omega_{ac}(q)$) the higher order emission processes Eq. (\ref{eq:encons-nph}) are also forbidden. Indeed, by representing the left hand side of Eq. (\ref{eq:encons-nph}) as 
the sum of $n$ terms 
\begin{eqnarray}
(\omega_{opt}(k)-\omega_{opt}(k-q_{1})+(\omega_{opt}(k-q_{1})-\omega_{opt}(k-q_{1}-q_{2}))+...
\nonumber\\
+(\omega_{opt}(k-q_{1}-q_{2}-...-q_{n-1})-\omega_{opt}(k-q_{1}-q_{2}-...-q_{n}))
\nonumber
\end{eqnarray}
and applying the condition of the lack of Cherenkov emission to each term as $(\omega_{opt}(k-q_{1}-q_{2}-...-q_{p-1})-\omega_{opt}(k-q_{1}-q_{2}-...-q_{p}))<\omega_{ac}(q_{p})$ we prove that the left hand side in Eq. (\ref{eq:encons-nph})  is always less than its right hand side. Thus, it is natural to expect that at zero (or sufficiently low) temperature no optical phonon decoherence takes place. 

At finite temperature the phonon scattering can always take place with the energy conservation. For instance the four-phonon scattering must satisfy the energy conservation law in the form 
\begin{eqnarray}
\omega_{opt}(k)+\omega_{ac}(q)=\omega_{opt}(k-p)+\omega_{ac}(q+p). 
\label{eq:encons-4phsc}
\end{eqnarray}
Eq. (\ref{eq:encons-4phsc}) can be satisfied for any $k$ in the thermodynamic limit of the infinite system. For example the backwards scattering with $p=2k$ occurs due to  Umklapp processes for the acoustic phonon wavevector $q=\pi/a-k$. However, these processes  are characterized by  rates that are smaller by the factor $\eta^2 \sim 0.01$ compared to the single phonon absorption or emission   (see Eq. (\ref{eq:AnhIntC})).  Therefore, they can be negligible  for realistic molecular lengths. At low temperature $k_{B}T < \hbar c/a$ they are also exponentially suppressed since the  Umklapp process requires excited acoustic phonons with the energy of order of $\hbar c/a$. 

Thus in spite of an inevitable decoherence in an infinite system  due to higher order anharmonic interactions at a finite temperature, this decoherence should be weak  in the case of a forbidden Cherenkov's emission, compared to the regime where this emission is allowed. Our numerical study reported in Sec. \ref{subsec:ExcSt} shows no decoherence up to the room temperature for a quite reasonable strength of an anharmonic interaction and a reasonable chain length of  $N=11$ due to a  discreteness  of a finite system energy spectrum  system (see  Refs. \cite{LoganWolynes90,LeitnerArnoldDiffusion97}). Higher order anharmonic interactions with solvent should inevitably destroy decoherence since discreteness is negligible for solvent. However, in the regime of a forbidden Cherenkov's emission, this decoherence should  occur at very long times, since it emerges as the outcome of a weak molecular solvent interaction and in a higher order with respect to the atomic displacements. 

\section{Numerical investigation of optical phonon  decoherence }
\label{Sec:Diag}


\subsection{Targets and methods of the numerical studies. }

\subsubsection{Targets}

In this  section we investigate   dynamics of  optical phonons numerically in a periodic chain of $N$ unit cells.  This study is complementary to the analytical consideration of Sec. \ref{subsec:Cher}. Our main target here is to verify the theoretical  prediction that  the optical  phonon experiences  decoherence  at zero temperature only if   Cherenkov emission is allowed  by the energy conservation Eq. (\ref{eq:CherEmisN}). We assume that the initial state ($t=0$) is composed by a single optical phonon with the given wavevector $k$ and evaluate the survival probability $P_{k}(t)$ for this optical phonon to remain in that state solving numerically the Schr\"odinger equation with the Hamiltonian of Eq. (\ref{eq:FullH}).

If decoherence, indeed, takes place, then the infinite time limit of the survival probability $P_{\infty}^k =P_{k}(t=\infty)$ vanishes for the infinite polymer chain $N\rightarrow \infty$,  while in the other case it should remain finite. We analyze these survival probabilities $P^{k}_{\infty}$ in Sec. \ref{subsec:popInf} for different wavevectors and system sizes and find their behaviors quite consistent with the Cherenkov emission criteria Eq. (\ref{eq:CherEmisN}) for the absence or presence of the decoherence. 

If the Cherenkov emission is allowed, the survival probabilities $P_{k}(t)$ decay with time to almost zero.  We analyze their time dependences and find typical   lifetimes of excited optical phonon states in Sec. \ref{subsec:DecRate}. The obtained relaxation times are then compared to the predictions of the analytical theory, Eq. (\ref{eq:FGRAns}).

We also consider  a  finite temperature regime in Sec. \ref{subsec:ExcSt} using the level statistics characterizing eigenstate properties. The absence or presence of decoherence is related to localization or delocalization of quantum states within the Hilbert space of harmonic states with given populations of all phonon states. The localization and delocalization in the Hilbert space can be conveniently  characterized using the levels statistics\cite{ShklovskiiShapiro93,OganesyanHuse07} that is examined. 

\subsubsection{Methods}

To solve numerically the Schr\"odinger equation we use the  eigenstates of the harmonic problem Eq. (\ref{eq:Ham0}) as the basis states. For the molecule composed by $N$ unit cells these states are  represented by the set of $N$ integer numbers.  The first number $n_{opt}$ indicates the state of the optical phonon  with $n_{opt}= 1, ...N$ corresponding to wavevectors $k=2\pi \tilde{n}/L$  ($\tilde{n}=n_{opt}-1$ if $n_{opt}\leq N/2+1$ or $n_{opt}-N-1$ otherwise). The sequence of other $N-1$ numbers represents the numbers of acoustic phonons in the states enumerated by numbers $1,2,...N-1$ similarly to those for optical phonons. The acoustic phonon normal mode with $k=0$ is excluded from the consideration since it corresponds to the motion of the chain as a whole that is not coupled to vibrations. 

The initial state of the system is chosen as the  excited optical phonon in the absence of excited acoustic phonons as for  absolute  zero temperature.  A nearly zero  temperature regime ($T\leq 10$K ) can be realized experimentally, for instance, for the molecules immersed in a liquid He \cite{Stewart83} or a solid matrix \cite{ab20PegsExp}.

A numerical diagonalization of the full Hamiltonian, Eq. (\ref{eq:FullH}), is not possible because of its infinitely large Hilbert space similarly to the Fermi-Pasta-Ulam problem \cite{ab19FPU}. However, since the states with a large number of acoustic phonons possess a very high energy compared to the energy of the state of interest (so they contribute only virtually), one can restrict the maximum number of acoustic phonons by the certain maximum number $n_{max}$,  following the approximations developed in Ref. \cite{ab19FPU}. Moreover, the contribution of those virtual states should decrease exponentially with increasing the number of excited acoustic phonons and therefore it should eventually become negligible when we include the sufficiently large number of acoustic phonons into consideration. 
The choice of $n_{max}$ is determined by a convergence of the numerical results with increasing $n_{max}$ (see e. g. Fig. \ref{fig:ConvPinf}). 

\subsection{Survival probability}
\label{subsec:survpr}

What can we  learn from the solution of the Schr\"odinger equation? From the perspective of ballistic transport it is important to know the probability for the optical phonon to remain in its initial state. This probability can be evaluated projecting eigenstates of the Hamiltonian to the states with the given optical phonon wavevector $k$ using the projection operator 
\begin{eqnarray}
\widehat{P}_{k}=\sum_{b}|k,b><b,k|, 
\label{eq:Projk}
\end{eqnarray}
where $b$ enumerates all states of acoustic phonons under consideration with the total number of phonons less or equal $n_{max}$. For the state described by the wavefunction $\Psi$, one can express the probability that the optical phonon has the wavevector $k$ as $P_{k}=<\Psi|\widehat{P}_{k}|\Psi>$.


It is convenient to express the probability $P_{k}$ in terms of the eigenfunctions,  $|\alpha>$, characterized by the  eigenenergies $E_{\alpha}=\hbar\omega_{\alpha}$, that can be expanded in a basis of the states $|b,p>$ with arbitrary wavevectors $p$ for optical phonons (cf. Eq. (\ref{eq:Projk})) as 
\begin{eqnarray}
|\alpha>=\sum_{b}c^{\alpha}_{pb}|p,b>.  
\label{eq:WfAlph}
\end{eqnarray}
Then the time-dependent wavefunction  with the initial condition of the absence of acoustic phonons (state $b=0$) and one optical phonon in the state $k$ can be expressed as
 \begin{eqnarray}
|\Psi(t)>=\sum_{\alpha}c_{0k}^{b*}c^{\alpha}_{pb}|p,b>e^{-i\omega_{\alpha}t}.  
\label{eq:WfTot}
\end{eqnarray}
Consequently, the probability $P_{k}(t)$ to find the optical phonon in its initial state $k$ can be expressed using the projection operator Eq. (\ref{eq:Projk}) as 
 \begin{eqnarray}
 P_{k}(t)=\sum_{\alpha,\beta}e^{i(\omega_{\beta}-\omega_{\alpha})t}c_{k0}^{\alpha *}c_{k0}^{\beta}\rho^{\beta\alpha}_{kk}, 
 \nonumber\\
 \rho^{\beta\alpha}_{kk}=\sum_{b}c^{\beta *}_{kb}c^{\alpha}_{kb},
\label{eq:Pt}
\end{eqnarray}
where $\rho^{\beta\alpha}_{kk}$  represents the  overlap  matrix for eigenstates $\alpha$ and $\beta$ with the fixed optical phonon wavevector $k$.


In the infinite time limit, the terms in Eq. (\ref{eq:Pt}) with $\alpha \neq \beta$ are averaged out due to oscillations and  only time independent terms survive. They  determine the infinite time limit of the probability that the optical phonon remains in the initial state $k$ as \cite{BerezinskiiGorkov79}
  \begin{eqnarray}
 P^{k}_{\infty}=\sum_{\alpha}|c_{k0}^{\alpha}|^2\rho^{\alpha\alpha}_{kk}.
\label{eq:Pinf}
\end{eqnarray}
The dependence of $P^{k}_{\infty}$ on the system size is different depending on whether the selected state is localized or delocalized within the Hilbert space. If it is delocalized, then $ P^{k}_{\infty}$ approaches zero in the thermodynamic limit of an infinite system, while in the localized regime it remains constant. The delocalization within the wavector space is equivalent to the decoherence, while localization suggests the ballistic transport to unrestricted  distances. 

Below we present the results for the survival probability in the infinite time limit (Sec. \ref{subsec:popInf}) and at finite times (Sec. \ref{subsec:DecRate}) and compare them to the Cherenkov emission criteria, Eq. (\ref{eq:CherEmisN}). The calculations are performed for the anharmonic interaction strength $V_{3}=0.2$ and maximum number of acoustic phonons $n_{max}=8$, as justified below. 

\begin{figure}
\includegraphics[scale=0.75]{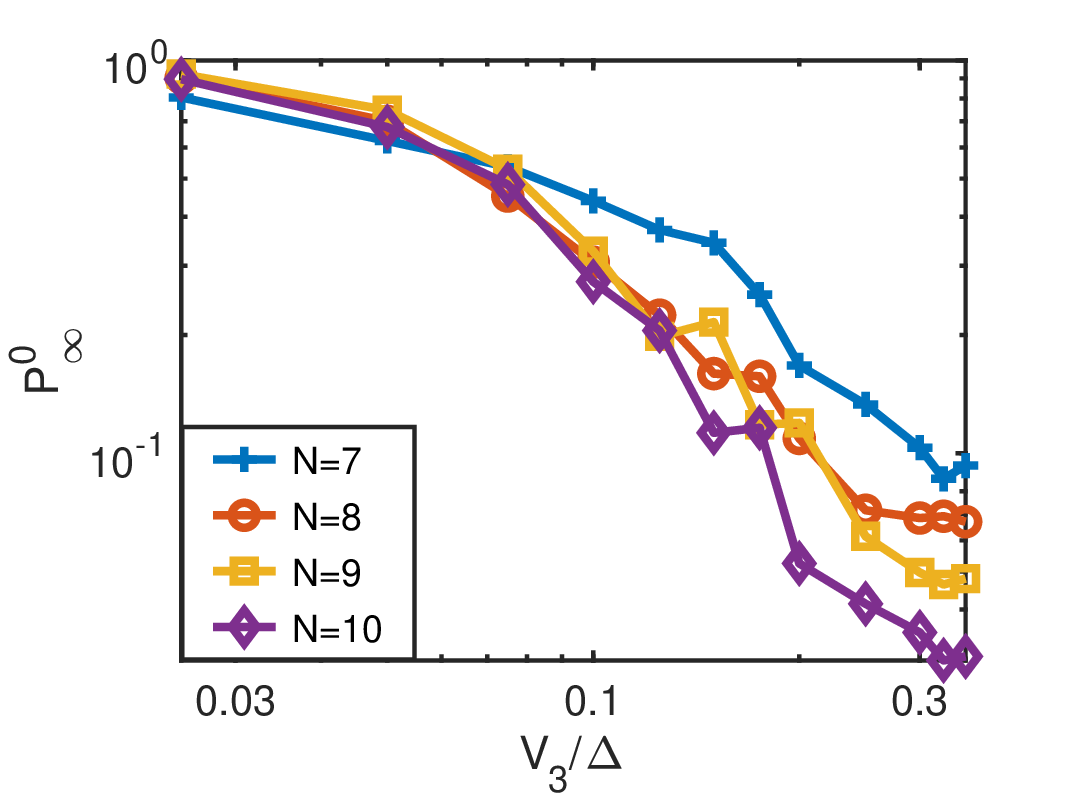} 
\caption{Dependence of the survival probability for the initial state with $k=0$ on the anharmonic  interaction strength. }
\label{fig:Vdep}
\end{figure}

\subsubsection{Choice of anharmonic interaction strength}
\label{subsub:V0.2}

We did most of the calculations for the anharmonic interaction strength $V_{3}=0.2\Delta$ in Eq. (\ref{eq:AnhIntC}). Being five times smaller than the optical bandwidth this interaction is quite realistic.  Particularly, it can be treated as a perturbation in the absence of Cherenkov's emission. In the presence of Cherenkov's emission for reasonably long molecules $N \geq 7$ this interaction leads to a decay of optical phonon accompanied by the emission of acoustic phonons.  


This is illustrated in Fig. \ref{fig:Vdep} showing the  infinite time survival probabilities for optical phonons with the  wavevector $k=0$  and indirect bandwidth, $\Delta<0$, so these states possess the maximum energy.  The dependence of survival probability on $V_{3}$ can be separated into two parts. At relatively small $V_{3}\leq 0.1\Delta$ this dependence looks like a straight line expressing a small correction to unity, proportional to $V_{3}^2$. The reduced value of $P_{\infty}$ for $N=7$ compared to  $N>7$ at small interaction $V_{3}$ is due to a  Fermi resonance for decay of the original optical phonon ($k=0$, the initial state $[1, 0, 0, 0, 0, 0, 0]$)  to the state with the wavevector $k=\pm 2\pi/L$ (the states $[2, 1, 0, 0, 0, 0, 0]$, and $[N, 1, 0, 0, 0, 0, 1]$), possessing a very close energy to that of the initial state. 
This behavior is typical for the regime of localization   \cite{BerezinskiiGorkov79}. 

At larger interaction constants $V_{3}>0.1\Delta$ a sort of chaotic behavior of a survival probability with changing  $V_{3}$ is seen, corresponding to decaying states delocalized in the Hilbert space of harmonic states. This regime always takes place for anharmonic interaction $V_{3}$ as high as $0.2\Delta$. For smaller $V_{3}$, longer chains are needed to observe a substantial decay of the initial state. The investigation of longer chains is difficult computationally since the size of the basis grows exponentially with the chain length. 

\begin{figure}
\includegraphics[scale=0.75]{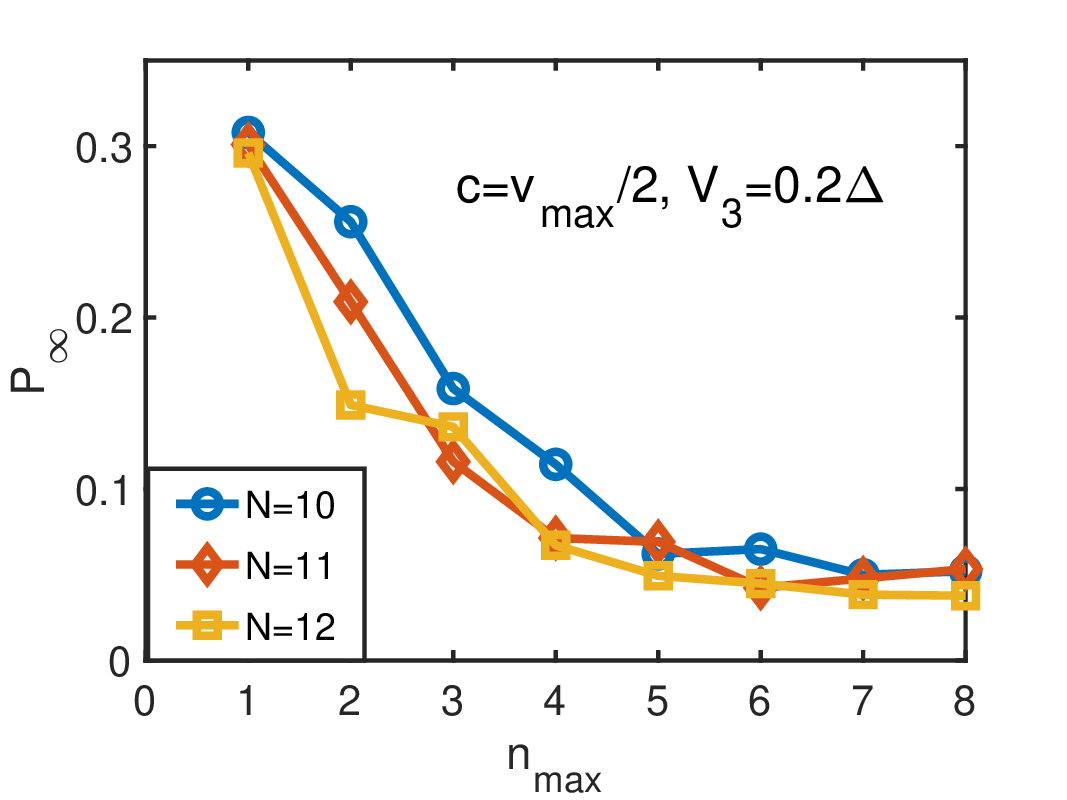} 
\caption{Size dependence of survival probability for the initial state with $k=0$, negative bandwidth and on the maximum number of acoustic phonons $n_{max}$ for the molecule with the substantial  Cherenkov's effect $v_{max}=2c$. }
\label{fig:ConvPinf}
\end{figure}

We do not consider interactions $V_{3}$ comparable or larger than the bandwidth since this is not very realistic. Indeed, anharmonic interactions should be smaller by around a factor of ten, compared to harmonic ones, Eq. (\ref{eq:AnhIntC}).   In addition, the strong anharmonic interaction  can substantially reduce the optical phonon bandwidth due to a polaron effect \cite{ab20Transv}. Consequently, Cherenkov's emission will be forbidden in that regime at low temperature. 

The speed of sound is considered within the domain $v_{max}/2\leq c \leq 2v_{max}$ since our target is to investigate the crossover between these two regimes, while the trends of interest  are expected to be strengthened far from the crossover emerging at $c=v_{max}$.

\subsubsection{Maximum number of acoustic phonons}
\label{subsub:nmax}

In Fig. \ref{fig:ConvPinf} we investigate  convergence  of survival probabilities  increasing the maximum number of acoustic phonons $n_{max}$ for different  numbers of sites   $N$. The results are given for the  interaction strength $V_{3}=0.2\Delta$ and the minimum speed of sound  $c=v_{max}/2$, where convergence is the worst.  The convergence is seen  at $n_{max} \approx 8$, corresponding to basis sizes between $20000$ and $40000$. The number of significant states, indeed, should be finite, since  total energy of all created acoustic phonons should not strongly exceed the optical phonon bandwidth representing  maximum energy, that the optical phonon can donate to acoustic phonons. Our calculations are mostly performed at $n_{max}=8$. Error bars shown in graphs given below indicate  the difference of the results between $n_{max}=7$ and $n_{max}=8$. 

\begin{figure}
\includegraphics[scale=0.75]{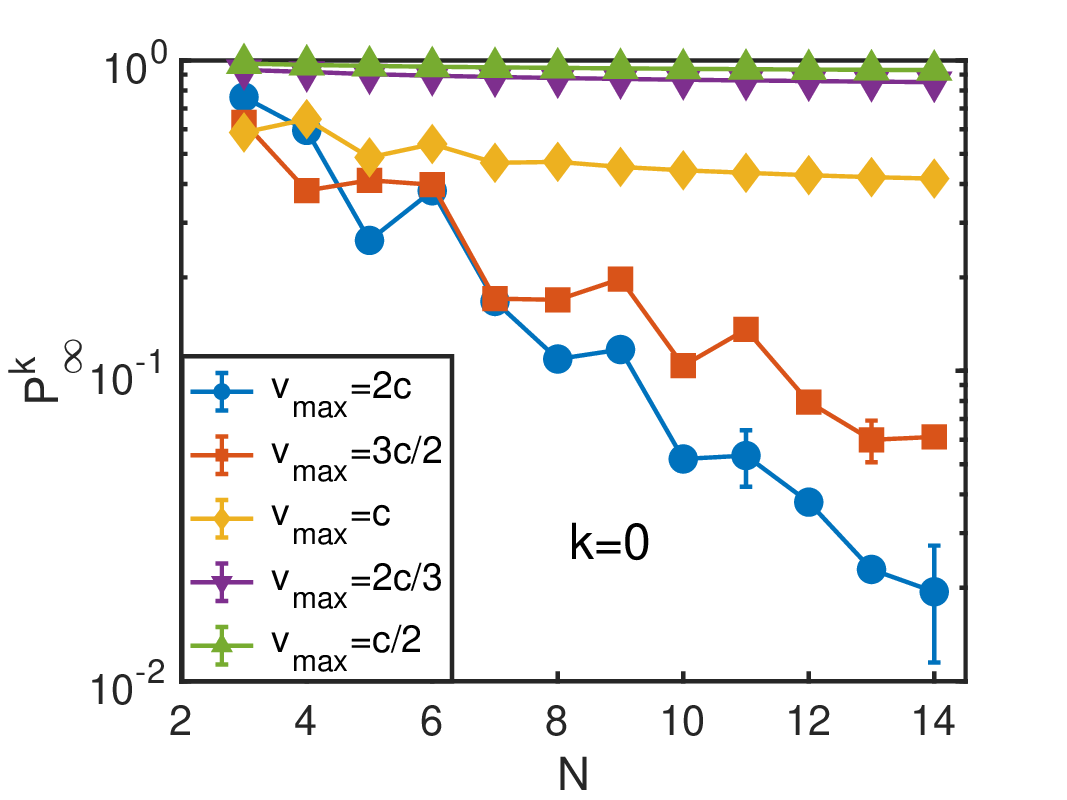} 
\caption{Size dependence of  survival probabilities   at maximum initial state energy ($k=0$ and indirect band $\Delta < 0$). }
\label{fig:InfMaxEninv}
\end{figure}

\subsubsection{Survival probability in the infinite time limit}
\label{subsec:popInf}

\begin{figure}
\includegraphics[scale=0.75]{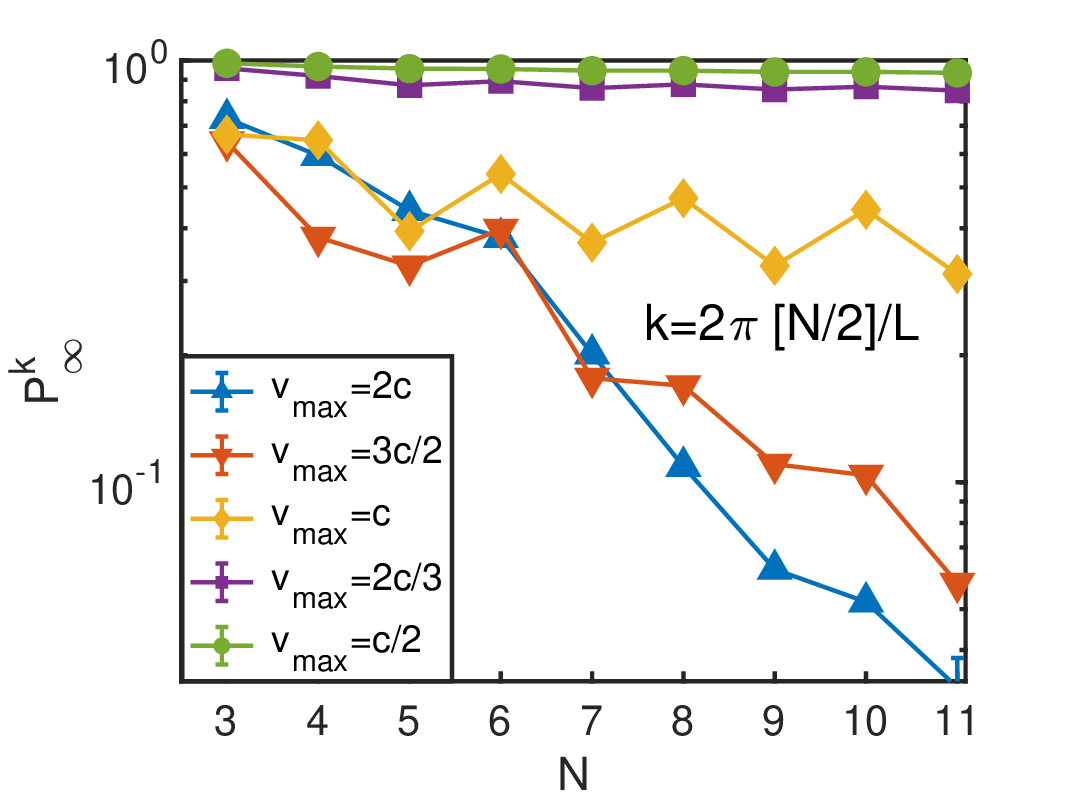} 
\caption{Size dependence of  survival probabilities  at a maximum initial state energy (initial state $k=2\pi [N/2]/L$ and  direct band $\Delta > 0$). }
\label{fig:InfMaxEn}
\end{figure}

Here we report dependence of the  infinite time survival probability  $P_{\infty}^{k}$  Eq. (\ref{eq:Pinf}) on the system size and the wavevector for different velocities of sound and chosen anharmonic interaction strength $V_{3}=0.2\Delta$ (see Sec. \ref{subsub:V0.2}).  Figs.  \ref{fig:InfMaxEninv} and \ref{fig:InfMaxEn} show survival probabilities for the optical phonon states possessing maximum energy vs a size of the molecule, i. e. the number of sites $N$. The sound velocity, $c$, changes from  $v_{max}/2$ to $2v_{max}$ Eq. (\ref{eq:OptPhgv}) around the expected threshold at  $c=v_{max}$ for the Cherenkov's emission.

Fig. \ref{fig:InfMaxEninv} represents an indirect  band ($\Delta<0$, Eq. (\ref{eq:OptPhgv})),  where the initially excited optical phonon possesses maximum energy and a wavevector $k=0$. The case of a direct band $\Delta>0$ is reported in Fig. \ref{fig:InfMaxEn}  for the initially excited optical phonon having a maximum energy and the wavevector  $2\pi [N/2]/L$, where the notation $[...]$ stands for the integer part of the argument in the square brackets and $L=Na$ is the molecular length. 

\begin{figure}
\includegraphics[scale=0.75]{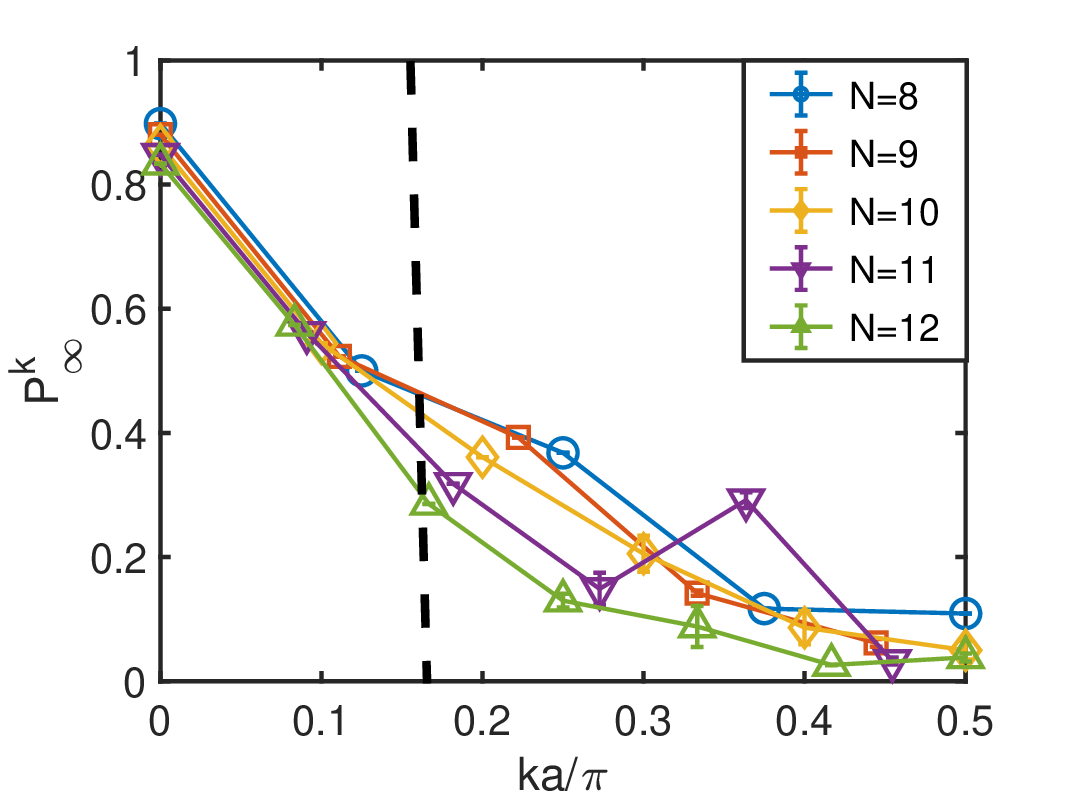} 
\caption{Dependence of  survival probabilities  on the initial state ($k$, $\Delta > 0$) in the regime of Cherenkov's emission $c=v_{max}/2$. }
\label{fig:Infk}
\end{figure}

Both graphs clearly show distinguishable dependences of survival probabilities $P_{\infty}^k$ on the number of sites $N$ indicating their decrease with $N$ for the allowed Cherenkov's emission ($c=v_{max}/2$ or $c=2v_{max}/3$) or its convergence to a constant of order of unity for the forbidden Cherenkov's emission ($c=3v_{max}/2$ or $c=2v_{max}$)  in full accord with our expectations. In the crossover regime $c=v_{max}$ survival probability decreases with $N$; yet the decrease is much slower compared to  smaller velocities of sound. Therefore, it is not clear whether  decoherence emerges in this regime or not. We leave this problem for future considerations. 


We also calculated survival probabilities for different initial wavevectors in the regime of allowed Cherenkov's emission  ($c=v_{max}/2$) as reported in Fig. \ref{fig:Infk}. The calculations were performed for the direct band ($\Delta>0$). The optical phonon group velocity depends on the wave vector as $v(k)=v_{max}\sin(ka)$ Eq. (\ref{eq:OptPhgv}). Consequently, at $\sin(ka)<1/2$ ($ka < \pi/6$) no decoherence is expected, while in the opposite regime of $\sin(ka)>1/2$ ($ka > \pi/6$) the substantial decoherence should be seen. The crossover between two regimes at $ka = \pi/6$is  indicated by the vertical dashed line in Fig. \ref{fig:Infk}.   These expectations are consistent with  Fig. \ref{fig:Infk}. 

\begin{figure}
\includegraphics[scale=0.75]{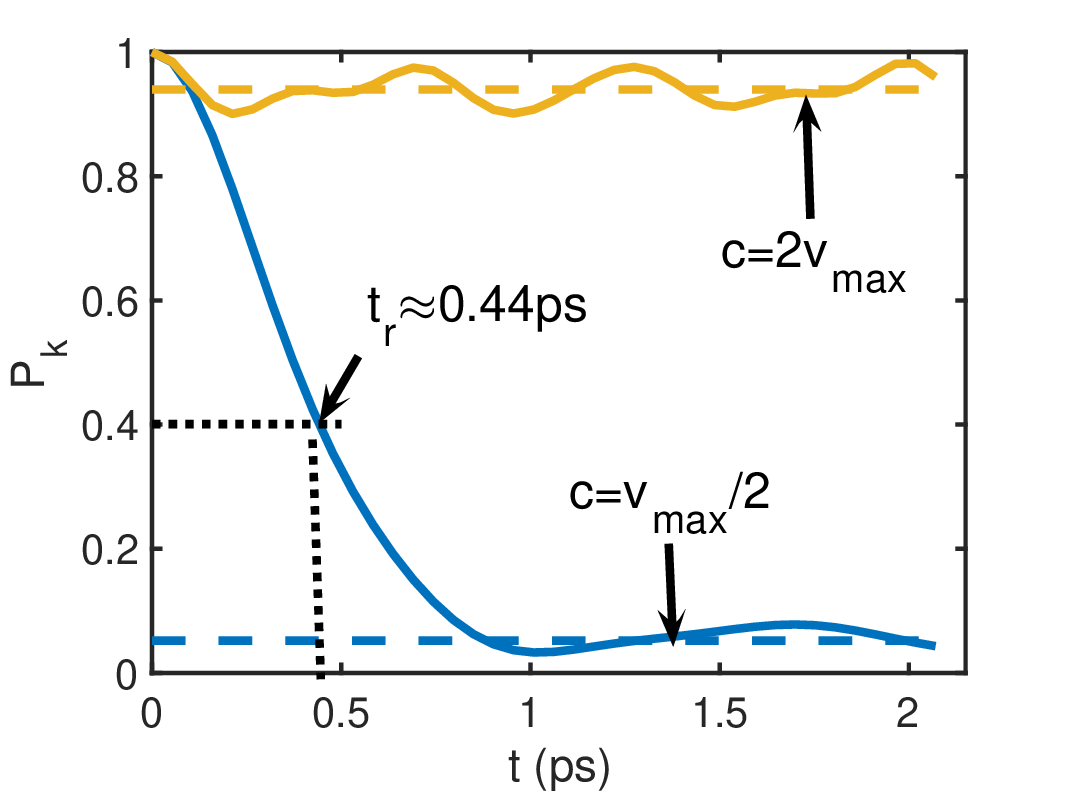} 
\caption{Time dependence of  survival probabilities  in the regimes of allowed and forbidden Cherenkov's emission for the initial optical phonon state with the maximum energy ($ka=\pi$ for $\Delta >0$) and the molecule composed by $N=10$ residues. Dashed lines show  infinite time limits of survival probabilities. Dotted lines illustrate a definition of the relaxation time (see text). }
\label{fig:Tdepk0N10}
\end{figure}

\subsubsection{Estimates of decoherence rates}
\label{subsec:DecRate}

Here we report the time-dependence of survival probabilities. Their time evolution is very sensitive to the specific phase of the system (allowed or forbidden Cherenkov's emission), as illustrated in Fig. \ref{fig:Tdepk0N10}. The time dependences of survival probabilities for the initial system state, chosen as a highest energy states of  optical phonon in the absence of acoustic phonons, are shown  there for the molecule with  $N=10$ sites. 
In the absence of the Cherenkov's emission ($c=2v_{max}$)  the survival probability shows coherent oscillations around its infinite time limit indicated by the dashed line with the amplitude comparable to the difference of initial survival probability $1$ and its infinite time limit $P_{\infty} \approx 1$. If the Cherenkov's emission is allowed ($c=v_{max}/2$), the survival probability,  $P(t)$, rapidly decreases with the time  converging to its infinite time limit $P_{\infty} \ll 1$. In this case  one can estimate  the phonon decoherence time as the time when the deviation of $P(t)$ from its infinite time limit gets  reduced by $e$. Consequently, we define the decoherence  time, $t_{r}$, as $P(t_{r})-P_{\infty}=(1-P_{\infty})e^{-1}$. This definition is illustrated by the dotted lines in Fig. \ref{fig:Tdepk0N10}. According to this definition we find $t_{r} \approx 0.44$ps.

\begin{figure}
\includegraphics[scale=0.75]{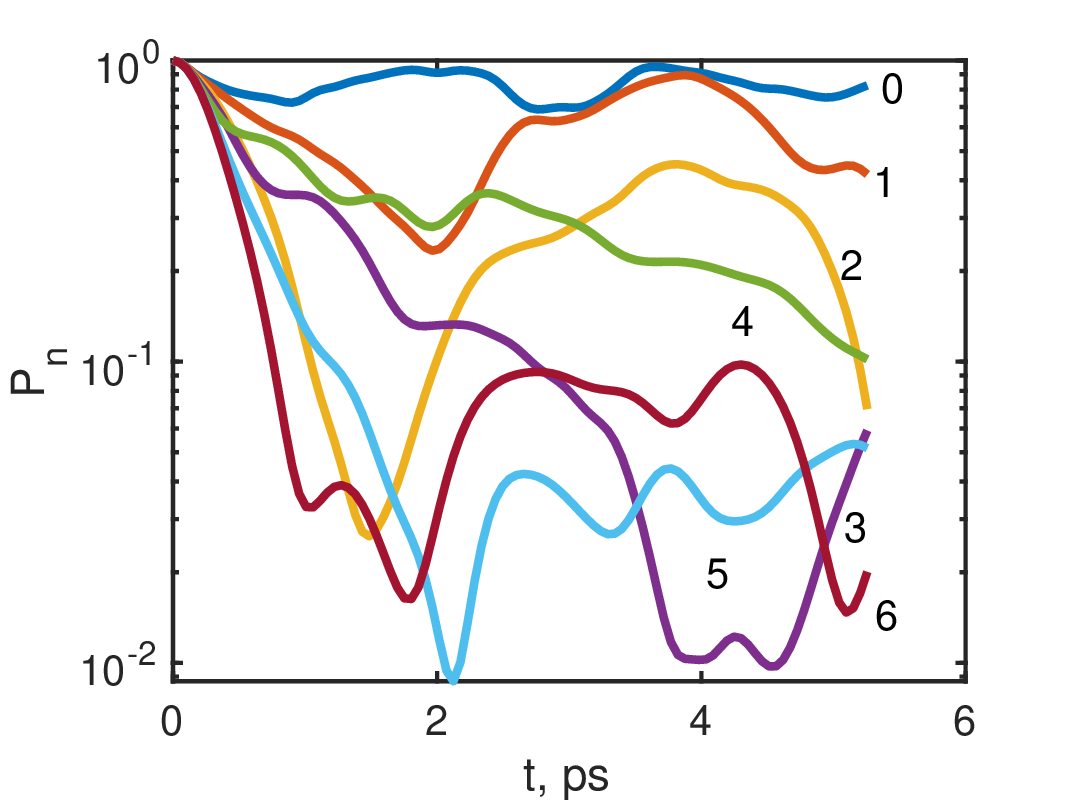} 
\caption{Time dependence of  survival probabilities  for allowed Cherenkov's emission ($c=v_{max}/2$, $N=12$, $n_{max}=6$). Numbers $n=0, 1.. 6$ assigned to each graph enumerates wavevectors of initially excited optical phonons, defined as $k=2\pi n/L$.}
\label{fig:TdepkN10}
\end{figure}

 In Fig. \ref{fig:TdepkN10} we show time evolution of the survival probabilities for the system of $N=12$ sites and all distinguishable wavevectors of the initially excited optical phonon ($n=kL/(2\pi)=0, 1,... 6$).  The direct band is considered ($\Delta>0$), so  optical phonon energy increases with increasing the wavevector, as in Fig. \ref{fig:Infk}. The relaxation of initial population is clearly seen  for four largest  wavevectors $n= 3, 4, 5, 6$ where the Cherenkov emission of acoustic phonons iis allowed, although it is slower for the second state $n=4$. The population remains of order of unity for two smallest wavevectors $n=0, 1$, where the Cherenkov emission is forbidden (cf. Eq. (\ref{eq:FGRAns})), which is consistent with their infinite time 
limits (see Fig. \ref{fig:Infk}). Low-amplitude oscillations (Fig. \ref{fig:TdepkN10}) with a quasi-period of a few picoseconds could be due to Fermi like resonances of the initial state and the states involving excited acoustic phonons.   The coupling of the states responsible for such period should be of order $1$cm$^{-1}$ and it can be induced by anharmonic interactions in the second or  third order or higher orders of perturbation theory with respect to it.  

The intermediate initial state $n=2$ is exactly at the crossover between two regimes. The transport velocity of the optical phonon for this state is equal to the speed of sound. The  state population shows large amplitude oscillations with the time. The analysis of its nature is beyond the scope of the present study.

\begin{figure}
\includegraphics[scale=0.75]{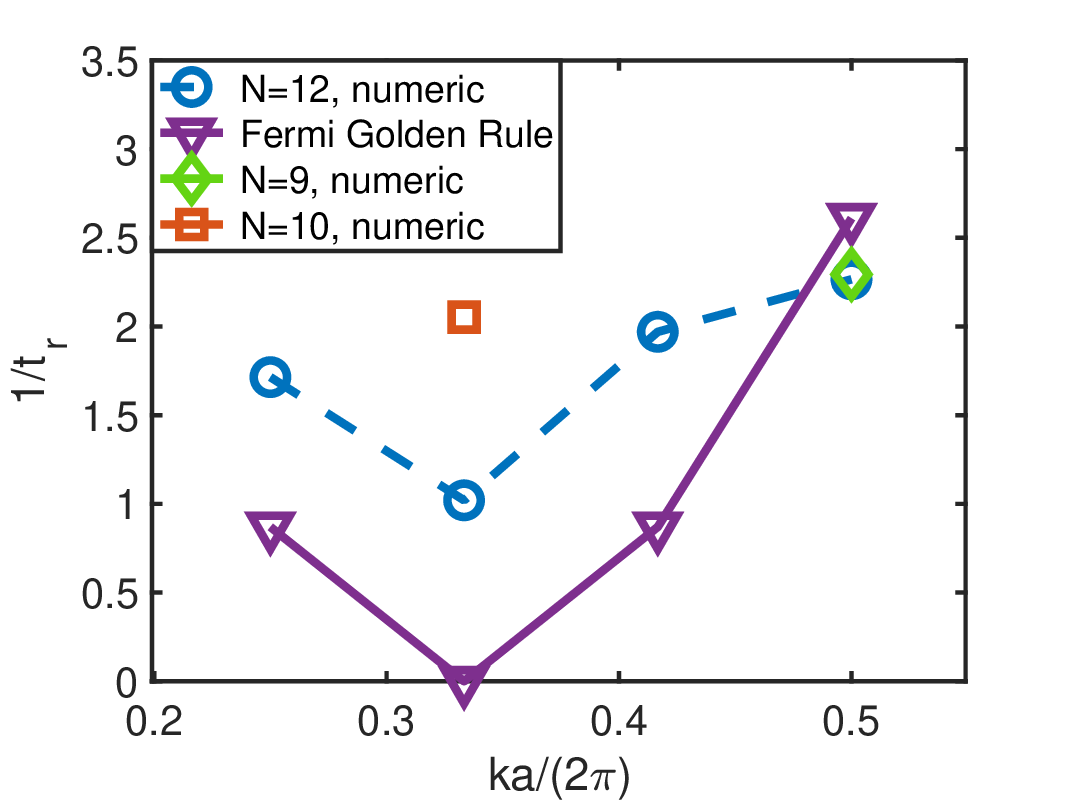} 
\caption{Estimates of inverse decay times using numerical solution  of the Schr\"odinger equation and Fermi Golden rule for different initial states. The data  for several other sizes are shown to examine a finite size effect.}
\label{fig:Relaxvsk}
\end{figure}

We estimated decay  times for all initial wavevectors corresponding to the delocalization regime $ka/(2\pi)=3, 4, 5, 6$ ($N=12$, $n_{max}=6$) and compared  them to the Fermi Golden rule estimates, Eq. (\ref{eq:FGRAns}), in Fig.  \ref{fig:Relaxvsk}, where they are expressed vs. wavevector dependent parameter  $ka/(2\pi)= (1, 2..N/2)/N$.  The results are consistent within the order of  magnitude except for the second state ($ka/(2\pi)=1/3$) from the top, where the Fermi Golden rule predicts a zero  relaxation rate. This anomalously  long time is due to a zero squared cosine factor in Eq. (\ref{eq:FGRAns}) for this specific state,  that is probably smeared out by the discreteness in the system under consideration and also can be originated from higher order processes in anharmonic interactions.  In the former case the decay rate should decrease with increasing the system size. Indeed, the decay rate of the initial state with the same wavevector for $N=9$ is twice faster, as shown in Fig. \ref{fig:Relaxvsk}, so the significant finite size effect is seen for this state. This is in  a sharp contrast to that for the wavevector $k = \pi/a$, where the comparison is reported  to the system with $N=10$ residues Fig. \ref{fig:Relaxvsk} and the difference of two relaxation times is less than $2$\%.  

In spite of the quantitative differences, the trends in  wavevector  dependence of decoherence times are consistent with each other for both approaches  at least in the domain of allowed Cherenkov's emission. We expect that the Fermi Golden rule should work better for larger system sizes,  although the corrections from the higher order  processes might be significant.  

\subsection{Localization and chaos in excited states}
\label{subsec:ExcSt}





As it was noticed before, the lack of decoherence of optical phonons in the absence of Cherenkov's emission $v_{max}<c$ suggests the localization of the system state within the Hilbert space of harmonic states with given mode population numbers \cite{LoganWolynes90,Leitner96,Srednicki94ETH,
LeitnerArnoldDiffusion97,BG2,BG1,BigwoodLeitner98}. Consequently, the persistent ballistic transport emerges since  optical phonon remains forever in its initial state with a finite probability. In the opposite regime with $v_{max}>c$,   decoherence should take place due to eigenstate delocalization in a Hilbert space of harmonic states. These expectations are  supported by numerical results in Sec. \ref{subsec:popInf}  for the zero temperature initial state with  no acoustic phonons. Since most of experiments are performed in the presence of acoustic phonons in the initial state, it is interesting to inspect whether this ballistic transport would persist there. 

According to our qualitative consideration of higher order processes in Sec. \ref{subsec:2HighOrd}, the  two phonon scattering of optical and acoustic phonons is always possible at a finite temperature in the infinite system.  This scattering  should lead to the decoherence in an infinite system.  Such scattering is a natural outcome of   anharmonic interaction, Eq. (\ref{eq:AnhIntC}), considered in the present work, as the second order process, involving virtual absorption of an  acoustic phonon with its subsequent reemission. Therefore,  decoherence should emerge in the system under consideration  at a finite temperature in the thermodynamic limit of an infinite system size.  However, in a finite system,  delocalization in the Hilbert space and, consequently, decoherence emerge  only at sufficiently high temperature \citep{LoganWolynes90,BG1,15LeitnerReview}. Below we  estimate the crossover temperature between these two regimes for the present model with anharmonic interaction $V_{3}=0.2\Delta$ and forbidden Cherenkov's emission $c=2v_{max}$ for the molecule containing $N=11$ sites. To distinguish localized and delocalized states we use level  statistics, as described below for the indirect band $\Delta<0$ and the states with total momentum $k=0$.

The localization or delocalization possesses different energy  level statistics. Delocalized ergodic states  are characterized by the Wigner-Dyson level statistics \cite{ShklovskiiShapiro93}, originated from the level repulsion. This repulsion results  in  a vanishing probability for a zero energy difference. Energies of localized states are characterized by the Poisson level statistics due to  their independence, because  localized states located far away from each other within the Hilbert space do not overlap. 

Level statistics can be conveniently represented by the averaged ratio of minimum to maximum differences between successive eigenenergies ($\Delta_{n}=E_{n+1}-E_{n}$) of the system \cite{OganesyanHuse07} 
\begin{align}
    \left< r \right> = \left< \frac{\min\left(\Delta_n,\Delta_{n+1}\right)}{\max\left(\Delta_n,\Delta_{n+1}\right)} \right>.
    \label{eq:lst} 
\end{align}
Localized states are characterized by $\left< r \right> \approx 0.3863$, while delocalized states are characterized by  $\left< r \right> \approx 0.5307$. To calculate the average value of  $<r>$  in a disordered system,  one has to find eigenvalues of the system Hamiltonian and average them over different realizations of disorder.

The $<r>$ parameter, Eq. (\ref{eq:lst}),  should be treated with caution in systems possessing integrals of motion different from the total energy, since energies of states having different values of those integrals   have no repulsion among  each other. In our system of interest the total quasi-wavevector is conserved. Therefore, we investigate separately the groups   of states with  total projections of the wavevector $k$ (remember that wavevectors different by an  integer number of $2\pi/a$ are considered as identical ones).  Also the  states  with $k=0$ (and $k=\pi/a$ for even $N$) possess an inversion symmetry, so we consider only symmetric states, while the results for antisymmetric states are identical to those of the symmetric states. 

\begin{figure}
\includegraphics[scale=0.4]{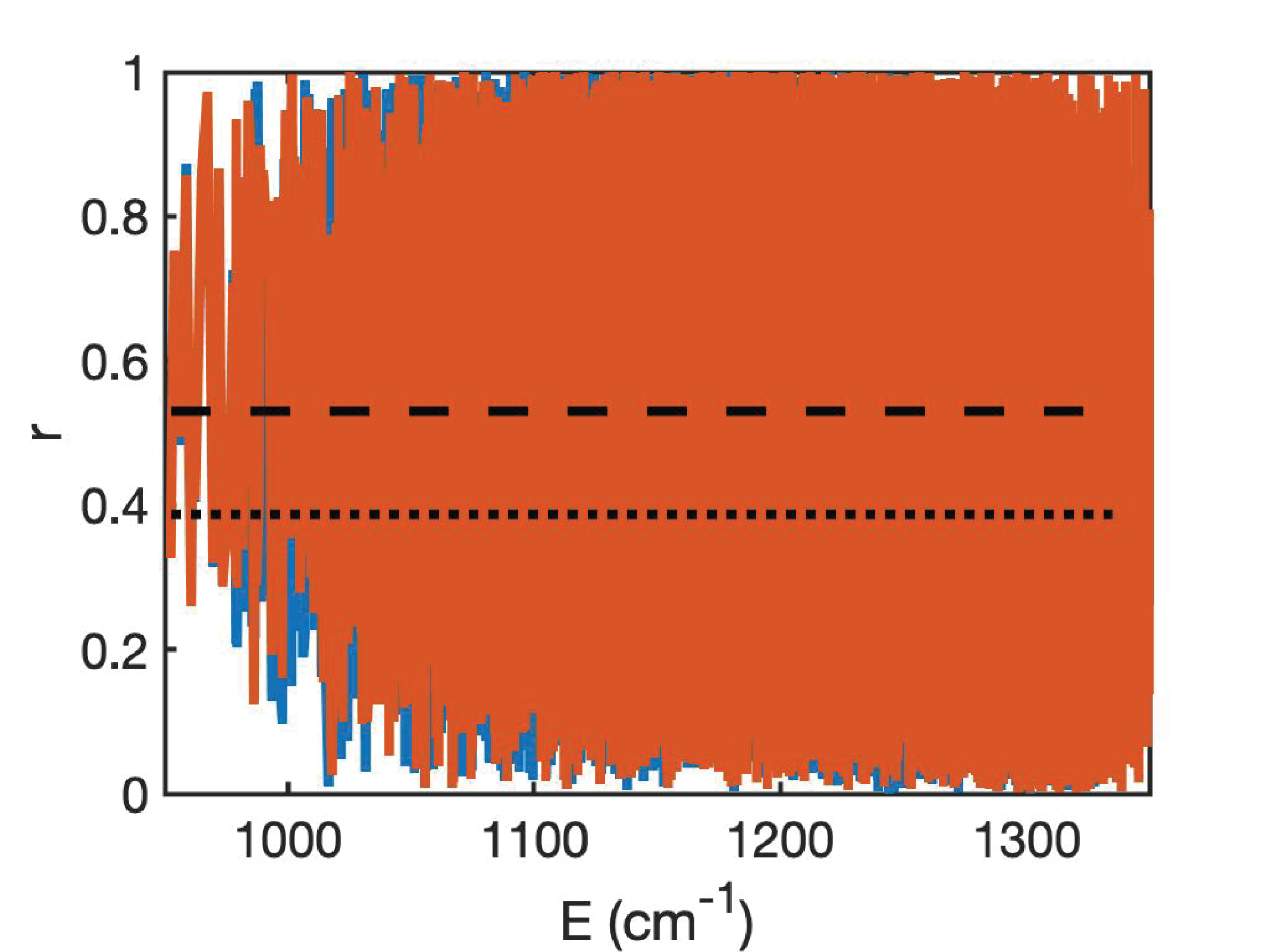} 
\caption{Dependence of the non-averaged ratio parameter Eq. (\ref{eq:lst}) on the energy for the Cherenkov's emission regime ($c=v_{max}/2$, $n_{max}=7$ and $8$ for  bottom and top graphs). Horizontal dotted and dashed lines indicate regimes of localization and delocalization (chaos).}
\label{fig:lstNotAv}
\end{figure}



\begin{figure}
\includegraphics[scale=0.75]{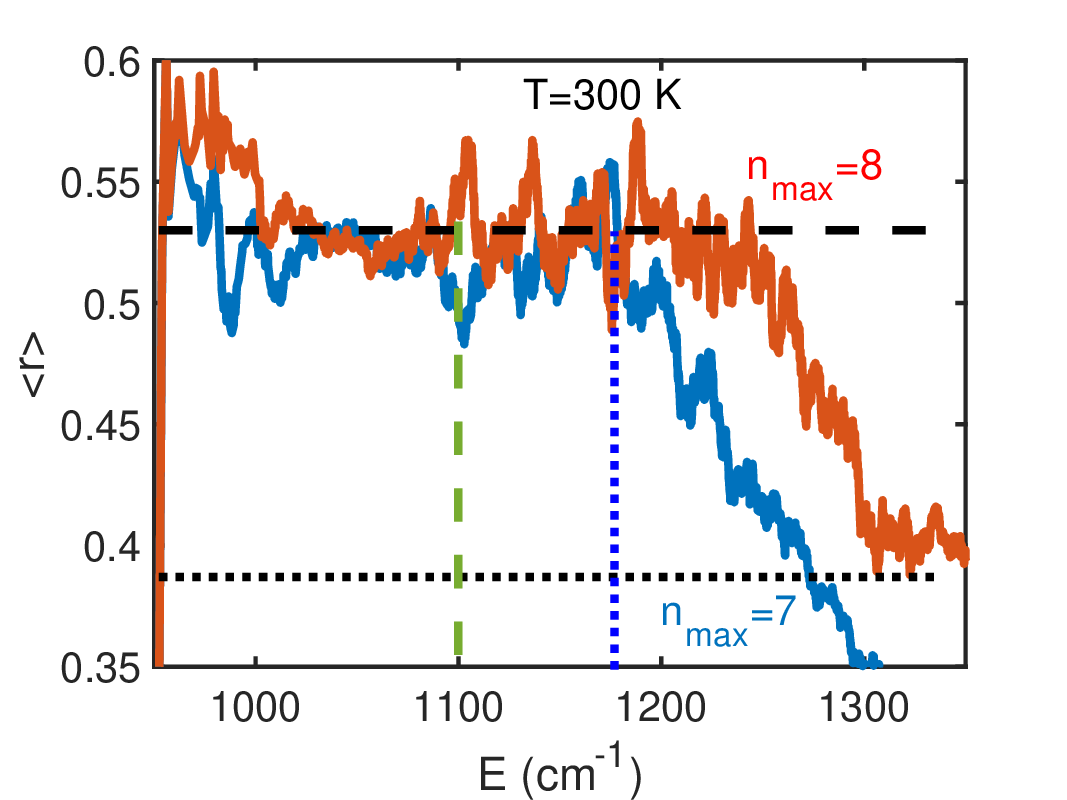} 
\caption{Dependence of the average ratio parameter $<r>$ Eq. (\ref{eq:lst}) on energy for the allowed Cherenkov's emission ($c=v_{max}/2$, $k=0$, $N=11$). Horizontal dotted and dashed lines indicate regimes of localization and delocalization (chaos).}
\label{fig:lstc05}
\end{figure}

There is no disorder in  the present problem. Without averaging, the ratio $r$ behaves as a random, strongly fluctuating variable, as shown in Fig. \ref{fig:lstNotAv}. The graph looks like the red spot (for $n_{max}=8$, in colors online) covering the blue spot (for $n_{max}=7$). To get rid of fluctuations we use the ratio $r$ averaged over $500$ ($n_{max}=7$) or $1000$ ($n_{max}=8$) adjacent states similarly to Ref. \cite{ab19FPU}. Then the average ratio parameter for the molecule  with allowed  Cherenkov's emission ($N=11$, $c=v_{max}/2$, $k=0$) shows a  clear signature of delocalization as depicted in Fig. \ref{fig:lstc05}. 

Remember that our consideration is approximate since the number of acoustic phonons is limited to a certain maximum number $n_{max}$. In the system under consideration with $N$ sites we cannot go beyond $n_{max}=8$ since at $n_{max}=9$ the Hilbert space grows above $50,000$ states making the calculations very slow.   The more acoustic phonons are included, the higher is maximum energy where the consideration is applicable. Comparing the results for $n_{max}=7$ and $n_{max}=8$ in Fig. \ref{fig:lstc05} we conclude that our analysis is applicable up to total energy of around $E_{max} \sim 1200$cm$^{-1}$. The average energy corresponding to the room temperature indicated by vertical dotted line is slightly below that energy, so the room temperature is covered by our study. The chaotic behavior is not surprising since delocalization of phonon states in the Hilbert space is expected for  allowed Cherenkov's emission. 

\begin{figure}
\includegraphics[scale=0.75]{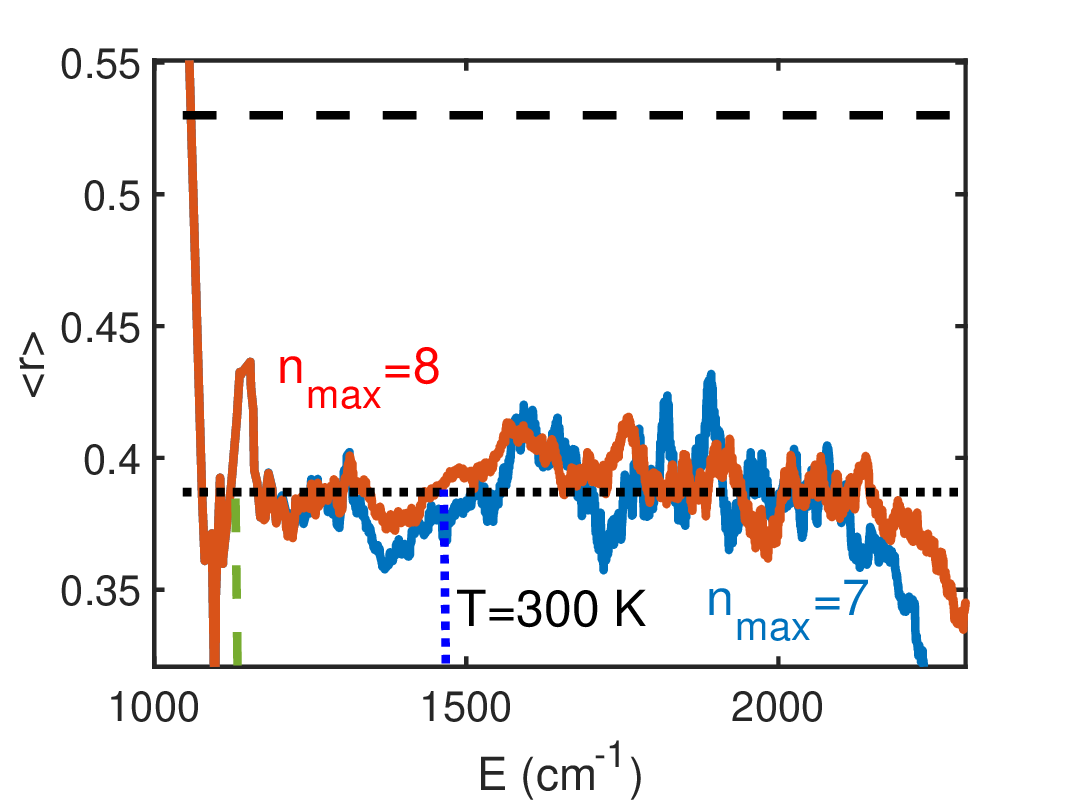} 
\caption{Dependence of average ratio parameter $<r>$ Eq. (\ref{eq:lst}) on energy for the forbidden Cherenkov's emission ($c=2v_{max}$, $k=0$, $N=11$).}
\label{fig:lstc2}
\end{figure}

Consider the opposite regime of forbidden Cherenkov's emission ($c=2v_{max}$). The average ratio $<r>$ dependence of energy is evaluated similarly to the previous consideration and reported in Fig. \ref{fig:lstc2}. Theory seems to be applicable to a very high energy up to $2000$cm$^{-1}$ corresponding to the temperature exceeding $500$K. At all energies under consideration the average ratio parameter is about $0.38$ corresponding to the Poisson statistics and localization \cite{OganesyanHuse07}. Consequently, no decoherence is expected in the present model for the parameters under consideration at experimentally relevant  temperatures.  

Thus we did not find the crossover to the delocalization regime at highest energy accessible in our calculations. Consequently, one can expect that  the localization in the Hilbert space and the lack of decoherence are robust for polyatomic molecules  in absence of Cherenkov's emission up to room temperature and even above it. Of course, other relaxation processes, including those involving the solvent, can also limit ballistic transport, in spite of the absence of Cherenkov's emission.  

The results for the level statistics and survival probability are approximately consistent  with each other for the energy $1100$cm$^{-1}$, which is the average energy of the  quantum state originated from  the initially excited optical phonon, considered in Sec. \ref{subsec:survpr}. As  indicated by the dashed vertical line, it shows the chaotic behavior in the regime of allowed Cherenkov's emission, as shown  in Fig. \ref{fig:lstc05}, and the localization for the forbidden Cherenkov emission \ref{fig:lstc2}. This consistency is not surprising.  Similar behavior was seen for interacting spin systems at an infinite temperature \cite{ab15MBL}, where the localization transition was examined using level statistics and spin-spin correlation function in the infinite time limit, which behaves similarly to the survival probability.

\section{Discussion of Experiments}
\label{sec:Exp}

Based on the previous consideration, it is natural to expect that the ballistic  transport of vibrational energy towards long distances can be realized only with the speed smaller than   speed of any sound propagating through the molecule.  Similar restrictions can be expected with respect to the sound propagating in solvent though the interaction with the solvent is weaker so the transport with the speed exceeding the speed of sound can carry to longer distances. Is this indeed true in the experiments 
\cite{Rubtsov2012pphynilultrfast,
Rubtsov2009Accounts2DIR,ab15ballistictranspexp,
ab19layla,ab19IgorReview} where the ballistic transport through optical  phonon bands has been observed? Below we compare the velocities of ballistic transport with  intramolecular and solvent speeds of sound. 

\begin{figure}
\includegraphics[scale=0.75]{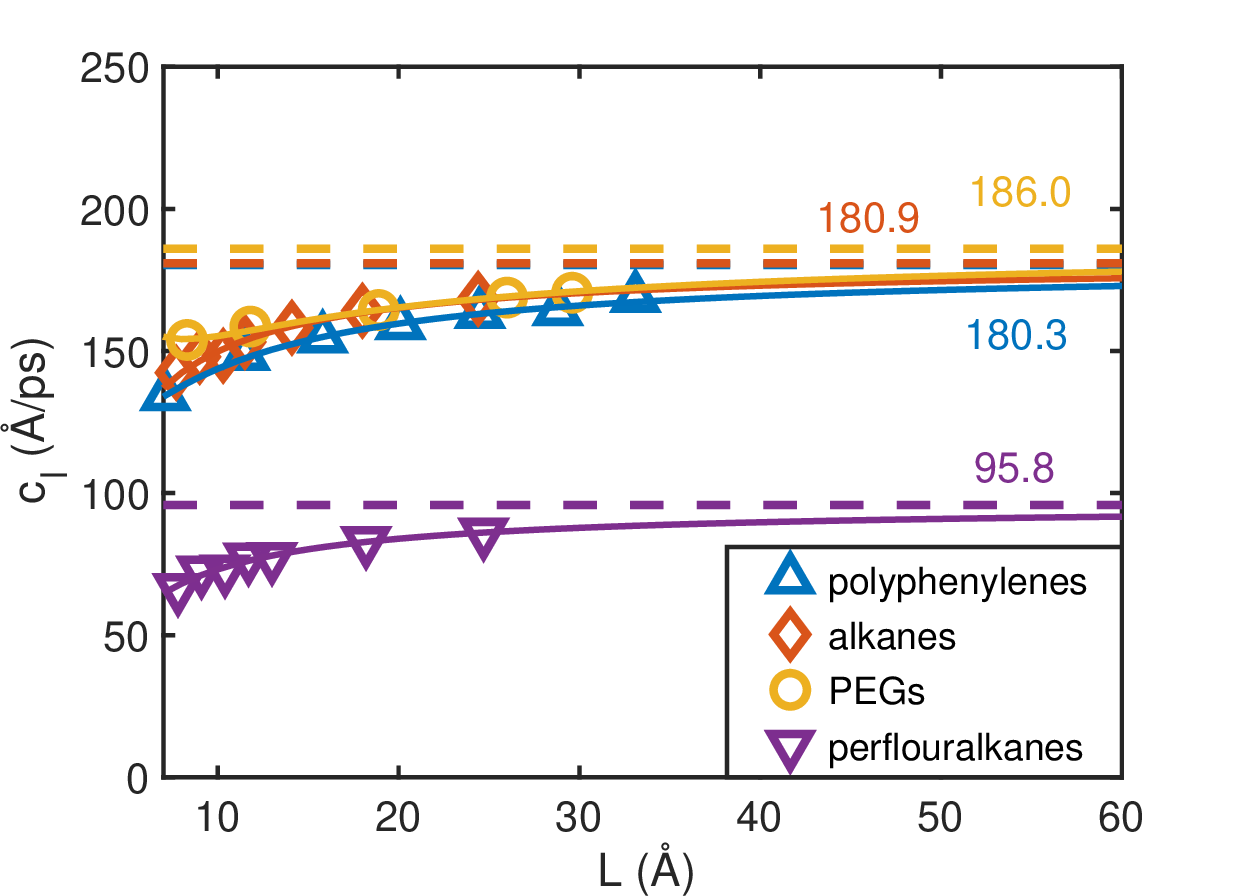} 
\caption{Computed longitudinal sound velocities as a function of the chain length for different chain types, modeled using Eq. (\ref{eq:c-exp}) (solid lines). The horizontal dashed lines and the numbers above them indicate the estimated sound velocities in the respective infinite chains.}
\label{fig:cl}
\end{figure}

\begin{figure}
\includegraphics[scale=0.75]{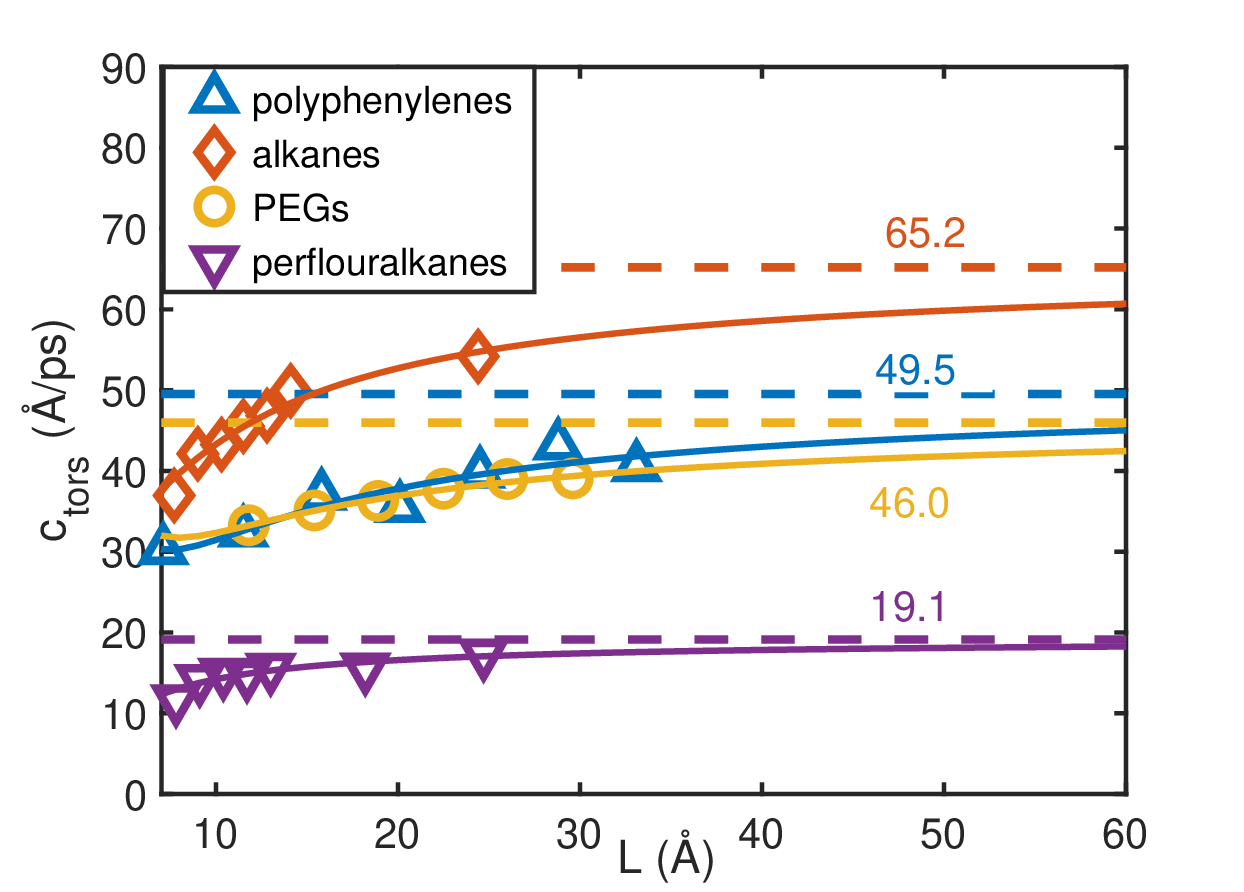} 
\caption{Computed torsional sound velocities as a function of the chain length for different chain types, modeled using Eq. (\ref{eq:c-exp}) (solid lines). The horizontal dashed lines and the numbers above them indicate the estimated sound velocities in the respective infinite chains.}
\label{fig:ct}
\end{figure}

Experimentally determined velocities of the optical phonon ballistic transport are   given in Table \ref{tab:Table1} and also shown in Figs. \ref{fig:cl}, \ref{fig:ct}. They should be compared to the speeds of sound in corresponding polymers and/or solvents. 

\begin{table}
\def\arraystretch{1}
\begin{tabular}{|c|c|c|c|c|}
\hline 
 Molecule & Oligophenylenes \cite{Rubtsov2012pphynilultrfast}& Alkanes  \cite{ab15JPCExpDec}  & PEGs \cite{ab19layla} & Perfluoralkanes  \cite{ab14PerFluoroAlkExp} \\ 
 \hline
$v_{exp}$ ({\AA}/ps)
 & 67 (67)& 14.7 (12) & 5.5 (4.5) & 3.8 (3.1)\\
\hline 
 $c_{l}$ ({\AA}/ps) & $180.3\pm 9.4$ & $180.9\pm 10.7$ & $186.0\pm 7.0$ & $95.8\pm 2.5$ \\
\hline 
$c_{tors}$ ({\AA}/ps) & $49.5\pm 1.0$ & $65.2\pm 1.2$ & $50.0\pm 4.5$ & $19.1\pm 0.2$ \\	
	\hline
Solvent & DMSO
& \multicolumn{3}{|c|}{CHCl$_3$}
\\
\hline
$c$ ({\AA}/ps) & 14.98 & \multicolumn{3}{|c|}{9.84} \\
 \hline
\end{tabular}
\caption { Comparison of optical phonon ballistic transport velocities,  found experimentally  ($v_{exp}$), with the speeds of longitudinal and torsional sound within the molecules ($c_{l}$, $c_{tors}$, respectively) and solvent ($c$). The experimental velocities are given as in Refs. \cite{Rubtsov2012pphynilultrfast,ab15JPCExpDec,ab19layla,
ab14PerFluoroAlkExp} for the through bond distance between the ends of periodic part of the molecule, while the actual velocities are given in parenthesis.}
\label{tab:Table1}
\end{table}

We did not find any available experimental or numerical data for the speeds of sound for either longitudinal or torsional modes in the oligomers of interest probed in the above-mentioned experiments. Therefore, we evaluated them using  DFT (B3LYP/6-311G++(d,p)) \cite{Gaussian}. To estimate the speed of sound we used size dependence of frequencies of the lowest energy modes for longitudinal or torsional vibrations. In a limit of a very long chain length $L$ they scale as $\omega_{l,t}(L) = c/L$, with the molecular length $L$ defined as the distance between the most far separated carbon atoms. This definition of the molecular length differs from Refs. \cite{ab14PerFluoroAlkExp,ab15ballistictranspexp,ab19layla} 
where the through bond distance was used. The experimental results for the transport velocities are given in table \ref{tab:Table1} following the original definition, while actual velocities are shown within the parenthesis. 

For each molecule we estimated the speed of sound using the expansion 
\begin{eqnarray}
c(L)=L\omega(L)=c+\frac{A}{L}+\frac{B}{L^2}, 
\label{eq:c-exp}
\end{eqnarray}
for the dependence $\omega(L)$ obtained numerically. 

The  fits of   size dependences of frequencies   using Eq. (\ref{eq:c-exp}) are illustrated in Figs. \ref{fig:cl}, \ref{fig:ct}. The results are presented in Table \ref{tab:Table1}.  Errors  were estimated comparing this fit with the fit by only two terms in equation (\ref{eq:c-exp}) with the parameter $B$ set to $0$. Since the difference between two approaches is relatively small it is reasonable to expect that higher order corrections are smaller at large system sizes used in our calculations.

For the vast majority of molecules investigated experimentally speed of sound values within the molecule exceed energy transport velocities given in the last row of Table \ref{tab:Table1} in agreement with our expectations.  The only exception for intramolecular sound waves emerges for  oligophenylenes \cite{Rubtsov2012pphynilultrfast}, where the speed of torsional sound is less than the energy  transport velocity. This  can be because only short oligophenylene molecules (up to three unit cells) were considered. 

The energy transport velocity in alkanes exceeds the speed of sound in solvent in spite of allowed Cherenkov's emission. This emission can occur much slower to the solvent compared to that within the molecule due to the substantially smaller coupling of vibrations to the solvent. 



\section{Conclusions}
\label{sec:Concl}

We found that the  velocity of ballistic energy transport through the optical phonon band of a periodic polymer chain is limited to the speed of sound. The faster transport is suppressed by the  Cherenkov - like emission of acoustic phonons, breaking down coherence during the time of  order of  $1$ ps due to  intramolecular interactions. 

Decoherence is substantially suppressed when the  Cherenkov's emission is forbidden. We found that the ballistic transport is persistent at zero temperature (in the absence of other relaxation processes, involving e. g. solvent).  To our surprise the lack of decoherence is  robust  up to relatively high temperatures exceeding room temperature for reasonably high anharmonic interactions.  Higher order anharmonic interactions with solvent should inevitably destroy decoherence  in the regime of a forbidden Cherenkov's emission; yet this decoherence should  occur at very long times. 

Our theoretical expectations are consistent with most of the  experimental observations where energy transport velocity  is, indeed, smaller than the intramolecular and solvent speeds of sound.

We ignored disordering that breaks down the momentum conservation  although it has a potential to violate Cherenkov's emission constraint. However, in the regime under consideration, where a localization length exceeds a molecular length, it should not be very significant; yet this is the subject for future studies.  

Following other work \cite{SegalNitzan03}, we also did not consider transverse acoustic phonons possessing the dispersion law different from that of sound waves, which can dramatically affect transport in the regime of strong anharmonic interactions \cite{ab20Transv}. The investigation of decoherence induced by absorption or emission of  transverse acoustic phonons will be investigated separately. 

\begin{acknowledgments}
This work is supported by the National Science Foundation (CHE-2201027).


\end{acknowledgments}

\section{Data Availability}

The data that support the findings of this study are available from the corresponding author upon reasonable request.
\bibliography{Vibr}
\end{document}